\documentclass[twocolumn]{aastex63}
\usepackage{amsmath,amssymb}
\usepackage{graphicx}

\shorttitle{Optical Spectrophotometry of Luhman 16}
\shortauthors{Heinze et al.}

\begin{document}

\title{Weather on Other Worlds. VI. Optical Spectrophotometry of Luhman 16B Reveals Large-amplitude Variations in the Alkali Lines}

\author{A. N. Heinze}
\affiliation{Institute for Astronomy, University of Hawaii, 2680 Woodlawn, Honolulu, HI, 96822, USA; aheinze@hawaii.edu}

\author{Stanimir Metchev}
\affiliation{Department of Physics and Astronomy, The University of Western Ontario, 1151 Richmond St, London, ON N6A 3K7, Canada; smetchev@uwo.ca}
\affiliation{Department of Physics and Astronomy, Stony Brook University, Stony Brook, NY 11794-3800, USA}

\author{Radostin Kurtev}
\affiliation{Departamento de Fisica y Astronomia,
Facultad de Ciencias, Universidad de Valparaiso, Av. Gran Bretana 1111, Casilla 5030, Valparaiso, Chile}
\affiliation{Millennium Institute of Astrophysics, Nuncio Monsenor Sotero Sanz 100, Of. 104, Providencia, Santiago, Chile}

\author{Michael Gillon}
\affiliation{Astrobiology Research Unit, Universit\'{e} de Li\`{e}ge, All\'{e}e du 6 Aout 19C, B-4000 Li\`{e}ge, Belgium}

\begin{abstract}
Using a novel wide-slit, multi-object approach with the GMOS spectrograph on the
8-meter Gemini South telescope, we have obtained precise time-series spectrophotometry of the binary brown dwarf Luhman 16 at optical wavelengths over two full nights. The B component of this binary system is known to be variable in the red optical and near-infrared with a period of 5 hr and an amplitude of 5--20\%. Our observations probe its spectrally-resolved variability in the 6000--10000\AA~range. At wavelengths affected by the extremely strong, broadened spectral lines of the neutral alkali metals (the potassium doublet centered near 7682\AA~and the sodium doublet at 5893\AA), we see photometric variations that differ strikingly from the those of the 8000--10000\AA~`red continuum' that dominates our detected flux. On UT 2014 February 24, these variations are anticorrelated with the red continuum, while on Feb 25 they have a large relative phase shift. The extent to which the wavelength-dependent photometric behavior diverges from that of the red continuum appears to correlate with the strength of the alkali absorption. We consider but ultimately reject models in which our observations are explained by lightning or auroral activity. A more likely cause is cloud-correlated, altitude-dependent variations in the gas-phase abundances of sodium and potassium, which are in chemical equilibrium with their chlorides in brown dwarf atmospheres. Clouds could influence these chemical equilibria by changing the atmospheric temperature profile and/or through cloud particles acting as chemical catalysts.
\end{abstract}


\section{Introduction} \label{sec:intro}

\subsection{Photometric Variability in Brown Dwarfs} \label{sec:photintro}

The 500K--2000K effective temperatures characteristic of L and T dwarfs enable the formation of condensate clouds of types not observable in our solar system --- including clouds of silicates and iron vapor. Such clouds have a profound influence on the spectra of brown dwarfs, and indeed there is a range of spectral types near the L-T transition \citep[roughly L8--T5;][]{Kirkpatrick2000} where the effective temperature is believed to be nearly constant and the spectral sequence is due to decreasing thickness of cloud cover as we move from the L dwarfs to the T dwarfs \citep{Ackerman2001}.

If clouds thick enough to affect the spectrum are distributed inhomogeneously in longitude over the atmosphere of a brown dwarf, rotationally modulated photometric and spectroscopic variability should result \citep[e.g.,][]{Artigau2009}. Studying such variability in detail can reveal the properties of the clouds. In particular, the variation in photometric amplitude as a function of wavelength can be diagnostic of the height of the cloud tops. In spectral regions with very low gas opacity (i.e., the bright continuum away from strong spectral absorption lines) regions with thick clouds extending to high altitudes are expected to be much darker than regions with thin, low-altitude clouds, because in the latter areas, the low gas opacity allows us to observe strong flux from deep, warm layers of the atmosphere. Hence, the less cloudy regions appear very bright in contrast to the dark, cloudy regions, and high-amplitude rotationally modulated variations are the result. By contrast, in spectral regions of high gas opacity (i.e., strong absorption lines), we cannot observe flux from deep in the atmosphere regardless of cloud conditions --- so the contrast between thicker and thinner clouds is expected to be much less and the photometric amplitude is expected to be lower. This is in some respects an oversimplified picture --- for example, the vertical temperature profile of the atmosphere (as well as the profiles of pressure and gas-phase elemental or molecular abundances) could be affected by the thickness of the clouds. Nevertheless, several observational results confirm the expectation of reduced photometric amplitude in regions of strong gas opacity --- in particular the near-infrared H$_2$O absorption from 1.35--1.50$\mu$m \citep[e.g.,][]{Apai2013,Buenzli2015a}.

\subsection{The Luhman 16 System} \label{sec:objintro}

The binary brown dwarf system Luhman 16 \citep[WISE J1049;][]{Luhman2013} is composed of two objects with spectral types  L7.5 and T0.5, respectively \citep{Burgasser2013,Kniazev2013}. With a distance of only 1.996 pc \citep{Bedin2017}, Luhman 16 is the third closest system to the Sun (after $\alpha$ Cen and Barnard's star) and its two components are the nearest known brown dwarfs. The system has an orbital period of $27.5 \pm 0.4$ years and a semimajor axis of $3.56 \pm 0.03$ AU, with dynamically determined masses of $33.5 \pm 0.3$ Jupiter masses for Luhman 16A and $28.6 \pm 0.3$ for Luhman 16B \citep{Lazorenko2018}.

Luhman 16B, the T0.5 secondary, is not only the closest T dwarf but also one of the most highly variable brown dwarfs known. \citet{Gillon2013} found it to vary with a period of about five hours and an amplitude exceeding 20\% in the optical on some nights. Among periodically varying brown dwarfs, this amplitude is exceeded only by the 26\% $J$-band variability detected by \citet{2M2139} in 2MASS J21392676+0220226, which was one of only two objects showing periodic variations with amplitudes greater than 10\% among more than 150 brown dwarfs monitored in surveys by \citet{Koen2013a}, \citet{Radigan2014}, and ourselves \citep{Metchev2015,Heinze2015}.

The very high photometric amplitudes observed for Luhman 16B suggest that its clouds are unusually large, dark, and inhomogeneous. They also mean Luhman 16B offers an opportunity to make measurements of wavelength-dependent amplitude variations with unusually high signal-to-noise ratio. Many studies have been made of the photometric and spectroscopic variability in this remarkable system: \citet{Gillon2013,Biller2013,Crossfield2014,Buenzli2015a,Buenzli2015b,Apai2021}, to list just a few. There is broad agreement that the rotation period of Luhman 16B is about $5 \pm 0.2$ hours, but finding a more precise value is difficult because the amplitude and light curve shape change dramatically on a timescale of just a few rotations \citep[e.g.,][]{Gillon2013,Apai2021}. The L7.5 primary Luhman 16A varies with an amplitude that is generally much smaller. This makes its rotation period more difficult to discern, but it has been estimated at about eight hours \citep{Mancini2015,Apai2021}.

Spectrally resolved studies of Luhman 16B's variability have found significantly reduced photometric amplitude in the near-infrared water absorption band from 1.35--1.50$\mu$m \citep{Buenzli2015a,comp}, consistent with the expectation discussed in Section \ref{sec:photintro} above. At the same time, \citet{Buenzli2015b} did {\em not} find evidence of an amplitude change in the iron hydride absorption band at 0.99 $\mu$m, which they interpreted as indicating that there are no completely unclouded regions in the atmosphere of Luhman 16B: the photometric variations are produced by variations in cloud thickness rather than by the complete absence of clouds in some areas. Gradations in cloud thickness as opposed to a stark cloudy vs. cloudless dichotomy would also be consistent with the results of \citet{Crossfield2014}, who used Doppler imaging based on high-resolution spectra in the 2.288--2.345$\mu$m regime to construct remarkable, disk-resolved maps of Luhman 16B. The maps show a total variation in photospheric surface brightness of only about 10\%. Large, coherent features would be required to sustain several-percent photometric variability with such a limited photospheric contrast range, and the \citet{Crossfield2014} maps show such features --- in particular, a kidney-shaped dark spot spanning about 70$^{\circ}$ in latitude and only slightly less in longitude.

We present a new spectrally resolved analysis of Luhman 16B, using Gemini/GMOS observations to probe shorter wavelengths than previous studies. In particular, our data cover the extremely broad, strong lines of the neutral alkali metals: the potassium doublet centered near 7682\AA~and the sodium doublet at 5893\AA. In Section \ref{sec:Data} we describe our observations and data reduction, and in Section \ref{sec:val} we validate our GMOS spectrophotometry by comparison with simultaneously obtained broadband imaging photometry. In Section \ref{sec:specrat}, we follow \citet{Apai2013}; \citet{Buenzli2015a,Buenzli2015b}; and \citet{comp} in exploring spectrally resolved variability by constructing ratios of spectra obtained in bright vs. faint photometric states. Section \ref{sec:anticor} describes the dramatic amplitude-reversal these spectral ratios reveal in the alkali lines, and in Section \ref{sec:custphot} we construct photometry in customized spectral bandpasses probing the time-resolved photometric behavior in these lines. We compare our results with expectations for simple cloud-driven variability in Section \ref{sec:expectation}, and consider but reject explanations based on lightning or aurorae in Section \ref{sec:emm}. We discuss more complex and realistic cloud-based explanations in Section \ref{sec:interp}, and offer our conclusions in Section \ref{sec:conc}.

\section{Observations and Data Processing} \label{sec:Data}

We observed Luhman 16 on the nights of UT 2014 February 24 and 25, using three different telescopes. We obtained optical spectrophotometry with Gemini South, infrared spectroscopy with Magellan, and broadband optical photometry with the TRAPPIST telescope \citep{TRAPPIST}. We present the two optical data sets here, having analyzed the Magellan infrared data in \citet{comp}.

\subsection{Observations} \label{sec:Obs}
We used the GMOS spectrograph \citep{GMOS} on the 8-meter Gemini South telescope in its multi-slit configuration to observe Luhman 16 simultaneously with 10 similar-brightness field stars. To avoid slit losses and obtain high-precision spectrophotometry, we used a very large, 5-arcsec width for our slits. To compensate for the significant sky background admitted through such wide slits, we employed the nod-and-shuffle mode of GMOS, a remarkable capability that enables very accurate sky subtraction even in the presence of strong fringing. In order to distinguish real spectral features from any column defects in the CCDs, we alternated pairs of images between two `spectral beams' by changing the grating tilt to move the spectra 50 pixels along the direction of dispersion. We will refer to these two grating tilt settings as beam 1 and beam 2. 

Due to the complexity of the nod-and-shuffle observations, the elapsed time for the acquisition of each GMOS science spectrum was about 480 seconds, but only 180 seconds consisted of actual integration on Luhman 16. This reduction in observing efficiency was more than compensated by the excellent sky subtraction enabled by the nod-and-shuffle mode. On February 24, we obtained twenty-six good quality science spectra spanning 7.1 hours of time, and on February 25 we obtained thirty-six good spectra spanning 8.2 hours. We used the OG515\_G0330 order-blocking filter on February 24, which blocks light shortward of 5150\AA. On February 25 we switched to the RG610\_G0331, which blocks light shortward of 6100\AA. This change was made out of concern that the spectra of comparison stars in the 10000--11000\AA~ region would be contaminated by their second order 5500--6000\AA~ spectra in the February 24 data, though such contamination would not directly affect the Luhman 16 spectra due to the object's extremely faint flux at blue wavelengths. Ultimately, the blue sensitivity of the February 24 spectra turned out to be more interesting than the theoretically more precise measurements in the 10000--11000\AA~ regime from the following night, because Luhman 16 did not exhibit any unusual behavior in the longer-wavelength regime.

The seeing was much better (0.69 arcsec) on Feb 24 than on Feb 25 (1.46 arcsec), but through deconvolution we were able to obtain resolved spectrophotometry of Luhman 16 A and B on both nights. The good seeing on Feb 24 enabled us to measure the separation between Luhman 16 A and B (perpendicular to the dispersion direction) at 7.738 pixels, or 1.131 arcsec. This precise measurement was important for deconvolving the spectral traces of the binary on both nights, as discussed below. 
Concurrently with our GMOS observations, we obtained unresolved photometry in the Sloan $z'$ filter using the TRAPPIST telescope \citep[see, e.g.,][]{TRAPPIST,Gillon2013}. The TRAPPIST data provide an important context for the GMOS spectrophotometry.

\subsection{Sky Subtraction} \label{sec:skysub}
\begin{figure*} 
\plottwo{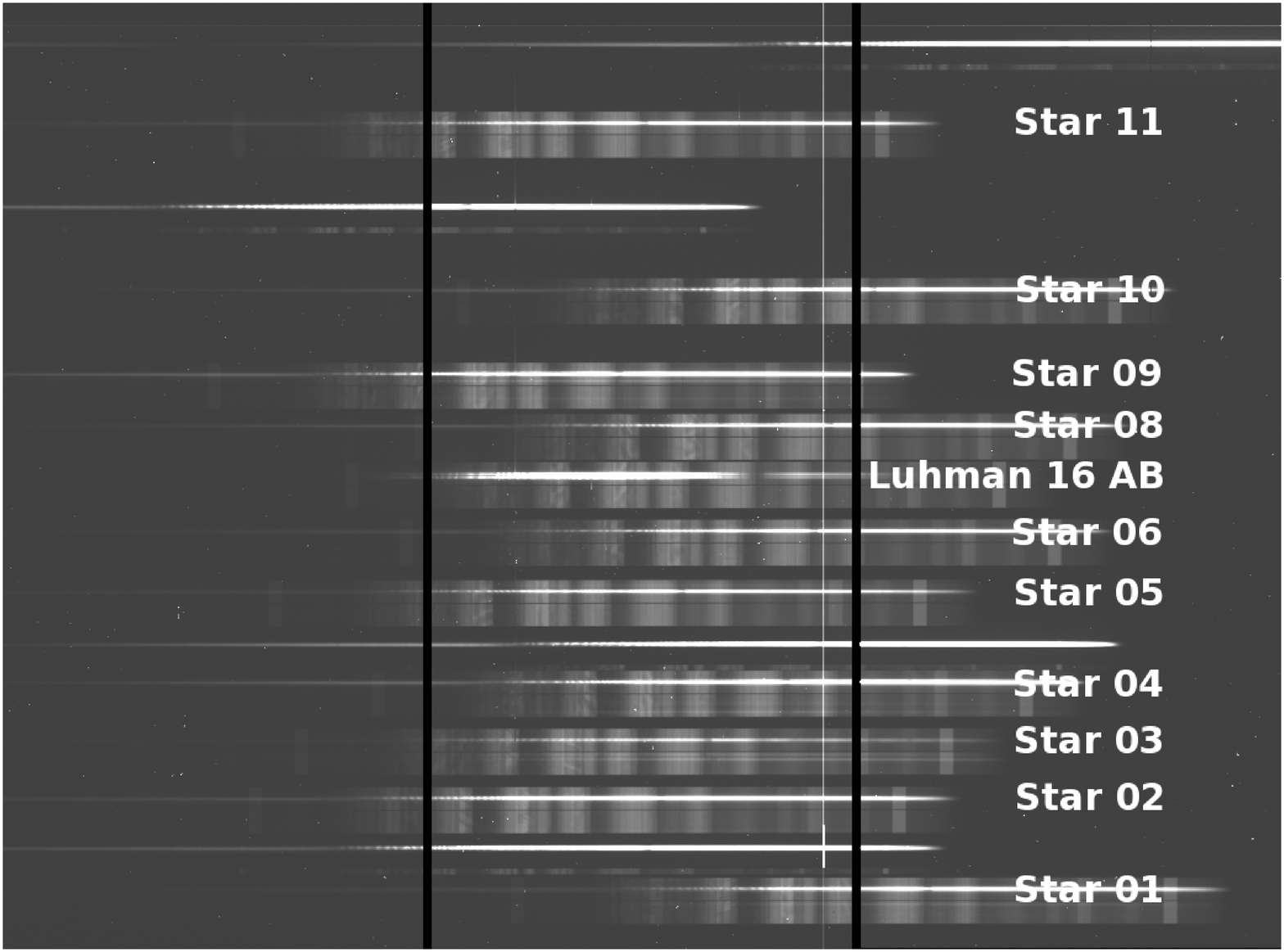}{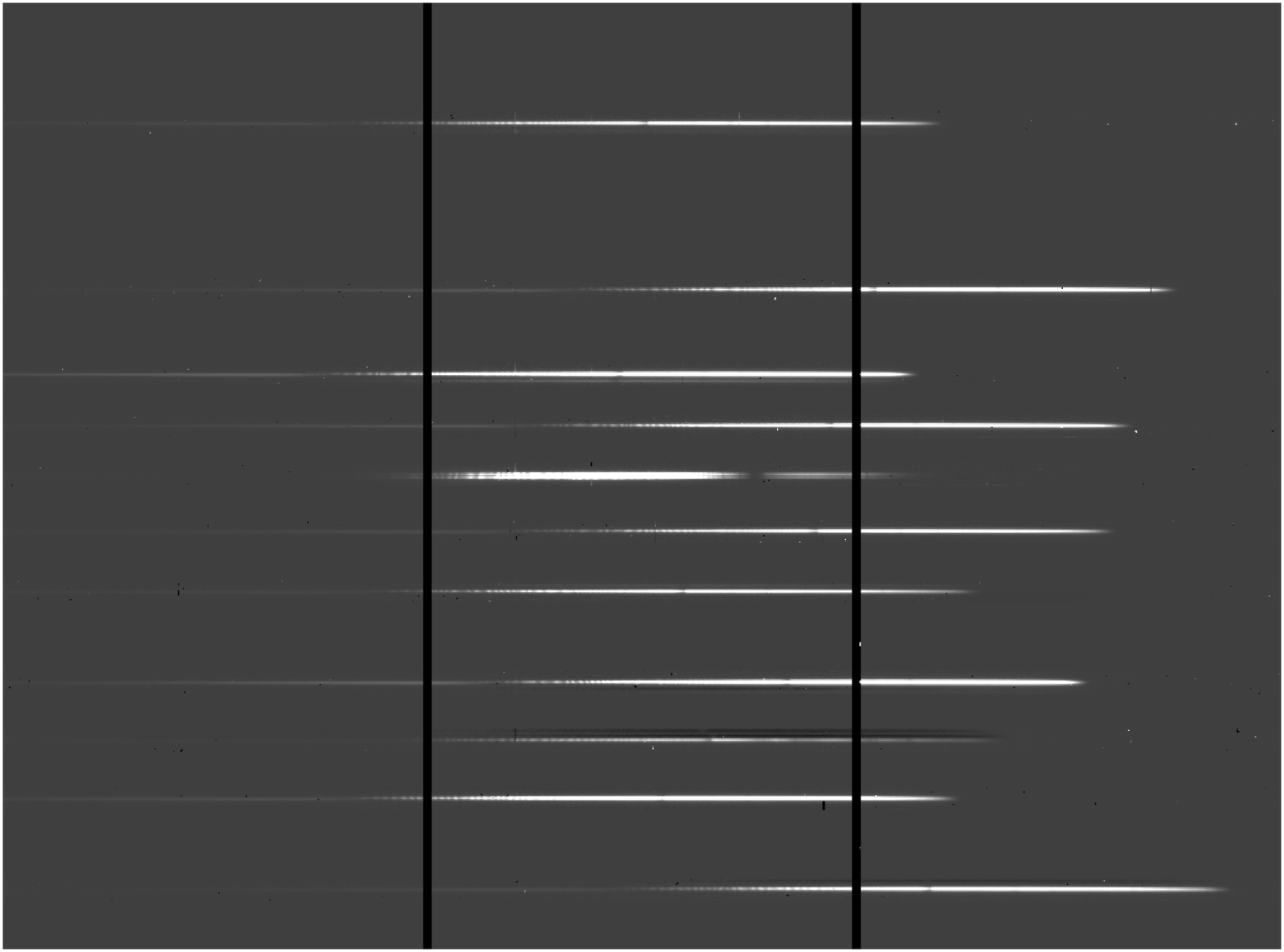}
\caption{Example of one of the GMOS science frames from which we extract spectrophotometric time series for Luhman 16 B.  \textbf{Left:} Raw, nod-and-shuffle image with Luhman 16 and ten similar-brightness field stars labeled.  Long wavelengths are toward the left. The starless spectra below each labeled star are not additional slits but rather the electronically shifted sky-only spectra produced by the nod-and-shuffle observing mode. The unlabeled bright spectra are pointing-check stars observed through smaller square apertures in the slit mask. \textbf{Right:} Sky-subtracted version of the same image. The sky background is completely removed, even at long wavelengths where it exhibits strong fringing. Star 03 becomes confused with negative spectra from fainter stars located in its nod-and-shuffle sky region, but no other object is affected this way.
\label{fig:rawdata}}
\end{figure*}

Since our images were taken using the nod-and-shuffle mode of GMOS, the science frames contain two spectra for each slit, offset vertically in the image by means of electronic `shuffling' on the CCD during the exposure.  One contains the target object and the other a sky background spectrum that is an extremely accurate match to the background that underlies the target spectrum. Thus, sky subtraction is a simple matter of shifting the sky spectrum vertically by an integer number of pixels and subtracting it from the science spectrum. We find that the nod-and-shuffle mode is very effective, and the complex, fringed sky background vanishes from the science spectrum essentially completely (Figure \ref{fig:rawdata}).

\subsection{Wavelength Calibration} \label{sec:arc}

\begin{figure*} 
\plottwo{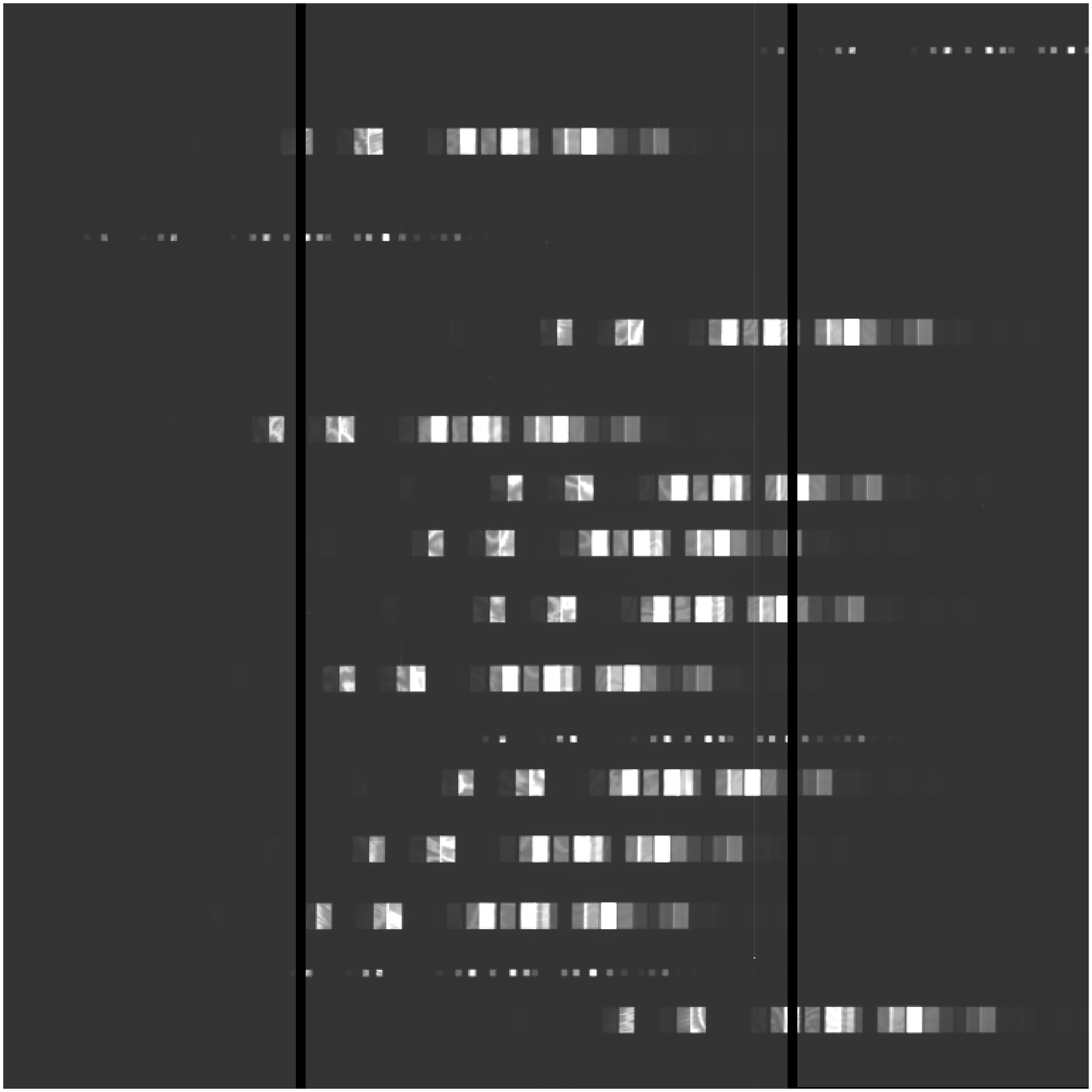}{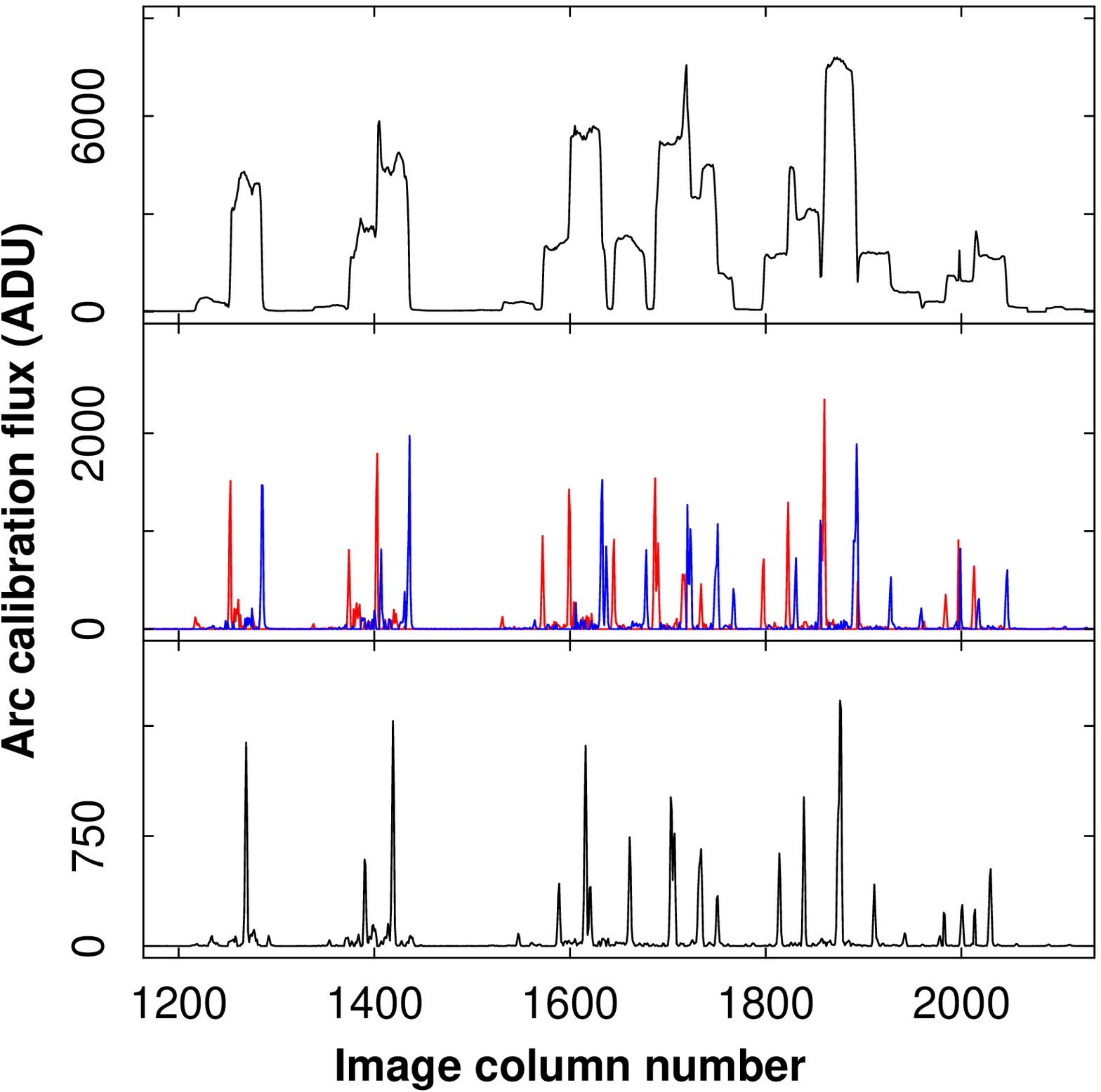}
\caption{Spectral calibration of our wide-slit data.  \textbf{Left Panel:} One of our arc calibration images, showing the wide, block-like slit images that correspond to spectral lines. \textbf{Right Panel, Top:} The result of summing over columns in the 7th slit on the image at left: the true arc spectrum convolved with the wide boxcar of the slit. \textbf{Middle:} The positive (red) and negative (blue) derivatives of the summed flux shown at top with respect to pixel position. \textbf{Bottom:} The average of the derivative vectors after they have been aligned by shifting each by half the slit width: thus, a deconvolved arc spectrum from which a precise wavelength calibration can be derived.
\label{fig:arc}}
\end{figure*}

We calibrate the wavelength solution using arc spectra.  Because of our wide, 5-arcsec slits, lines in the arc spectra appear as wide blocks (Figure \ref{fig:arc}).  We deconvolve these spectra using an edge-detection methodology, which begins with summing along image columns within the band of rows corresponding to each slit, and creating two 1-D vectors containing the positive and negative derivatives of the column-sums as a function of x pixel position. We discard all negative values of the derivatives, thus the positive derivative is nonzero only at the left-hand edges of `slit blocks' and the negative derivative is nonzero only on the right-hand edges of the blocks.  We find the cross-correlation peak of the two derivative vectors, which corresponds to the true slit width (5 arcseconds, or about 37 pixels). Finally, we align the two derivative vectors by shifting each by an amount equal to half the slit width, and average them to obtain a final deconvolved arc spectrum as shown in the bottom right panel of Figure \ref{fig:arc}.

We obtain a spectral wavelength solution by fitting the x-centroids of lines in the deconvolved spectrum vector as a quadratic function of the known wavelengths. We reject lines that show large residuals due to cosmic rays, bad columns, or proximity to the gaps in the detectors.  As 9--15 lines survive unrejected for each slit, the quadratic fit to wavelength remains well-constrained. The flux from Luhman 16 is bright enough for variability analysis from 6000--10000\AA, and the spectral pixel scale varies from 3.5 to 3.6\AA~per pixel over this range. In our Feb 24 data, when we used an order-blocking filter with a bluer cutoff wavelength of 5150\AA, Luhman 16's extremely faint `blue' flux is detected all the way down to the cutoff. The actual spectral resolution element depends on the seeing, but is about 25 \AA~under typical conditions: thus, our spectral resolution varies from $\mathrm{R} = 200$ to $\mathrm{R} = 400$.

\subsection{Spectrophotometry Overview} \label{sec:extract}

We extract both unresolved and resolved spectra from our images. The extraction methods are quite different, so in some cases the two types of spectra can offer quasi-independent confirmation that observed spectrophotometric behavior is intrinsic to Luhman 16 and not an artifact of our data processing. 

The unresolved spectra contain light from both components of the binary, while in the resolved extraction we use a novel form of deconvolution to obtain distinct spectra for Luhman 16A and Luhman 16B. In both cases, our goal is to obtain precise, spectrally resolved relative photometry. We are not primarily interested in producing a physically calibrated spectral energy distribution, although we do attempt such calibration in an approximate sense in Section \ref{sec:physcal}.

In processing the unresolved spectrophotometry, our reference for removing photometric effects imposed by Earth's atmosphere is an average over the simultaneously-observed reference stars that occupy the other slits in our MOS mask (after confirming the reference stars have no significant astrophysical variability). In the resolved photometry, we target Luhman 16B, and our reference for removing photometric effects from Earth's atmosphere is Luhman 16A --- hence, we make the assumption (which can later be checked) that the latter object is not significantly varying.

The advantage of the unresolved spectrophotometry is its simplicity. Its disadvantages are that the variations of Luhman 16B are diluted by the light from its less-variable companion, and that obtaining accurate spectrophotometry is problematic at wavelengths longer than 10000\AA~due to the faintness of the reference stars' spectra at these wavelengths and (in the February 24 data) their contamination by short wavelength flux from the stars' second-order spectra.

The advantages of the resolved spectrophotometry are the opportunity to study the two objects independently; the exquisite removal of telluric effects by probing two objects of identical color observed in the very same slit; and the ability to probe all the way out to the red limit of the GMOS CCDs at 11000\AA. Its disadvantages are the challenge of reliably deconvolving the blended spectra, and the possibility that the reference object Luhman 16A could itself be slightly variable --- although at most wavelengths we ultimately avoid the latter problem by solving algebraically for the photometric variations of both objects (Equation \ref{eq:solveAB}).

\subsection{Extracting Unresolved Spectra} \label{sec:extunres}

\begin{figure} 
\includegraphics[scale=0.5]{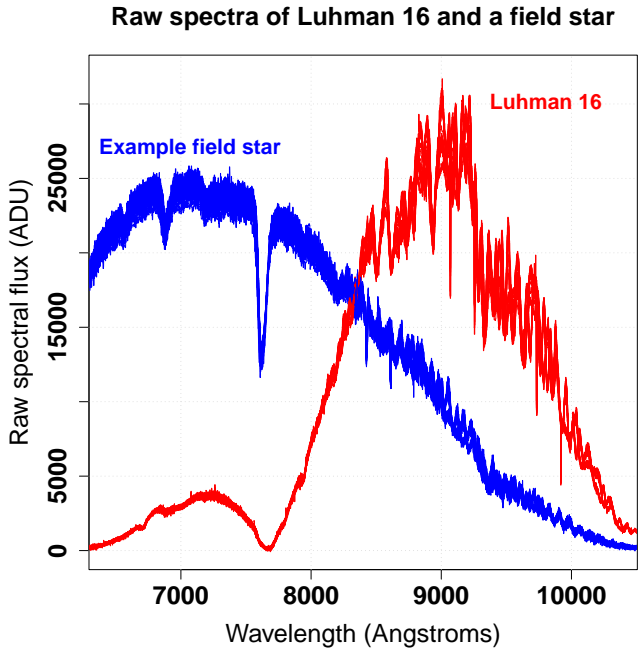}
\caption{Unresolved spectra for Luhman 16 (red) and a typical field star, from 34 different science images taken on the night of UT Feb 25. The field star is fainter than Luhman 16, but has been scaled up by a factor of three for easier comparison. The lack of large variation in spectral normalization testifies to the good transparency and stability of the Earth's atmosphere during our observations.
\label{fig:rawspec}}
\end{figure}

We extract unresolved spectra by simply summing along columns in each slit. Figure \ref{fig:rawspec} shows spectra from 34 different images taken on UT Feb 25.  The narrow width of the bands formed by the overplotted spectra illustrates the relatively small change in raw total flux through the whole night: atmospheric conditions were consistently photometric.

\begin{figure*} 
\plottwo{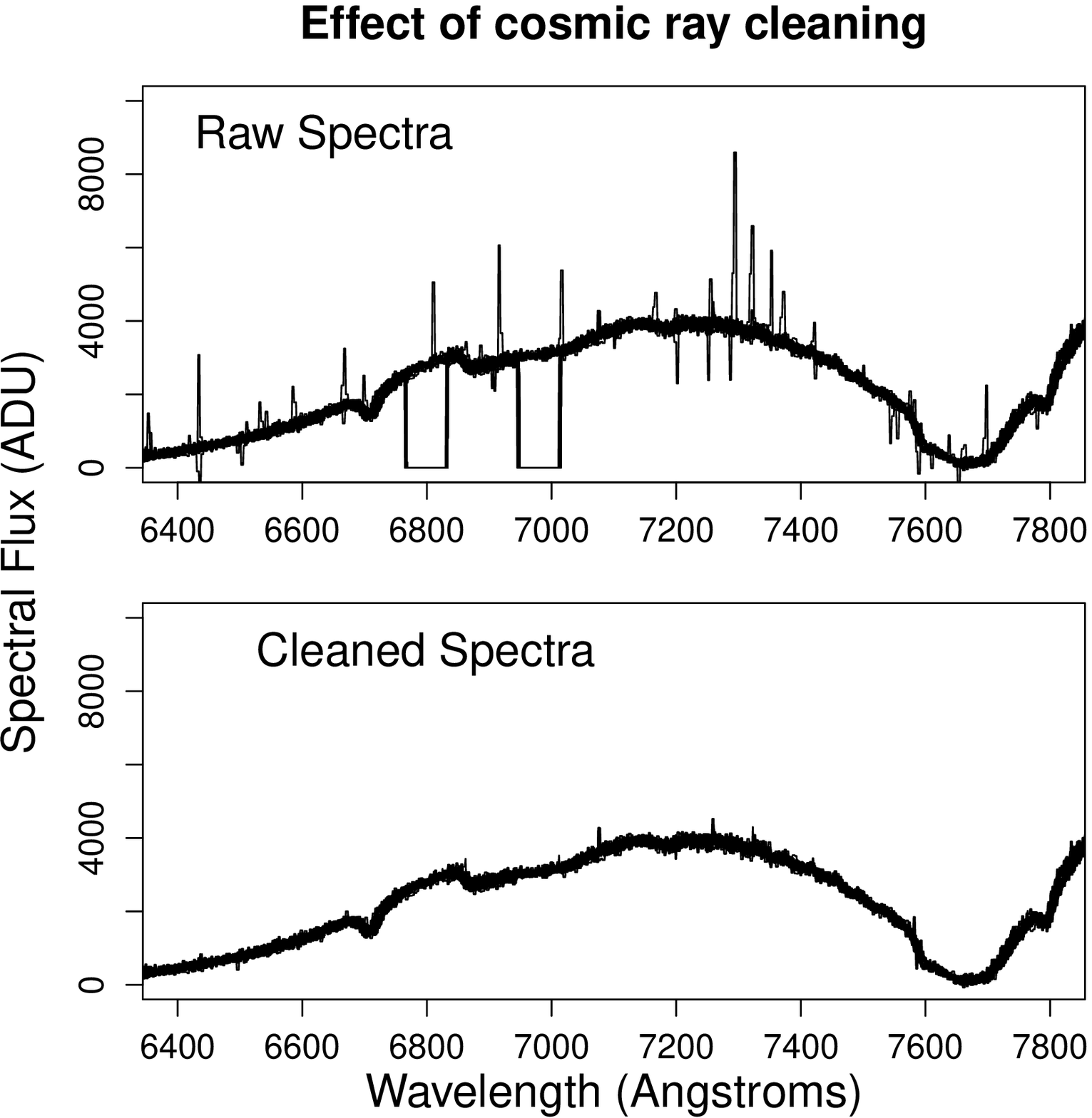}{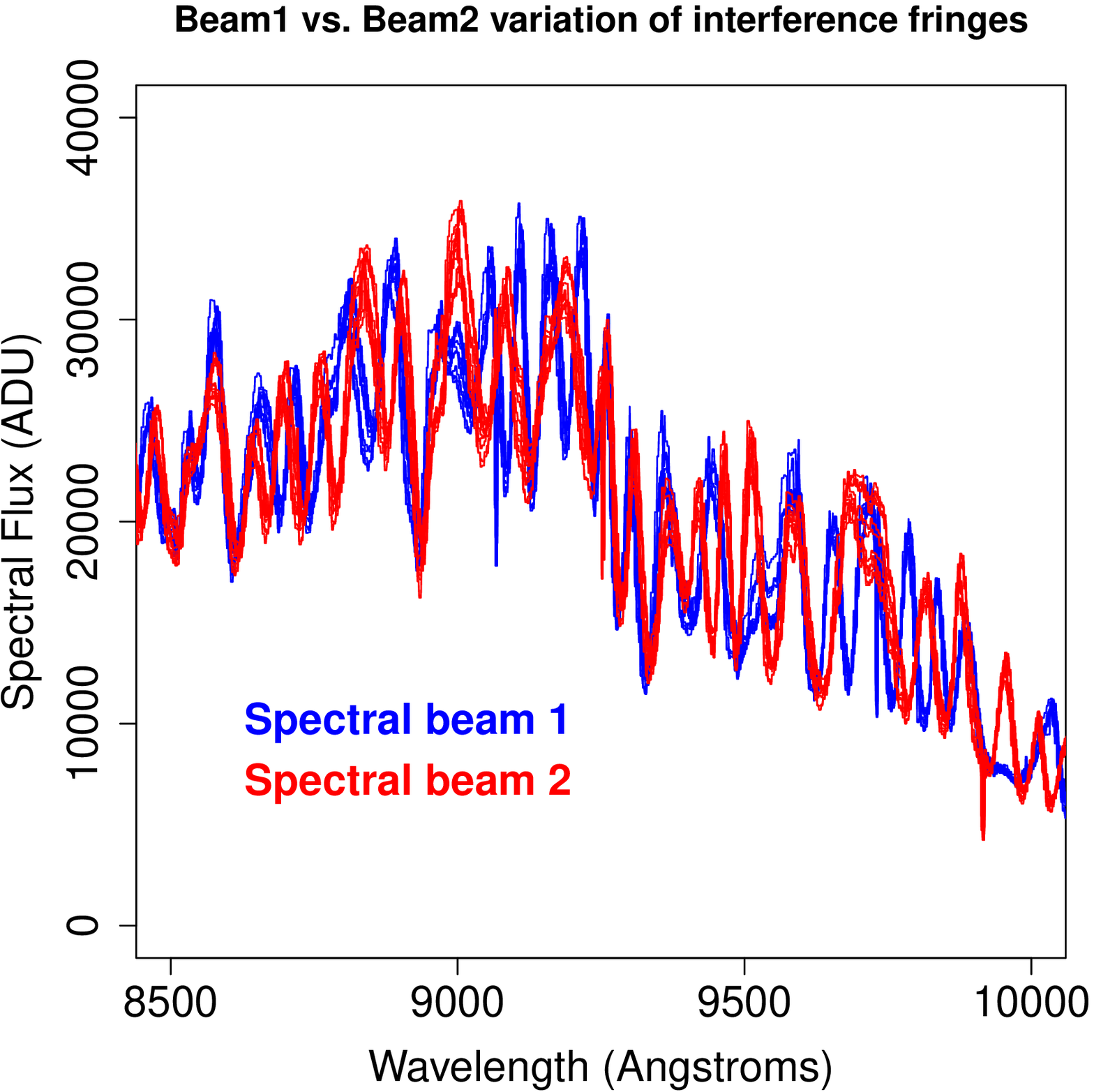}
\caption{\textbf{Left:} Result of cosmic-ray cleaning. All 26 unresolved spectra of Luhman 16 that were obtained on UT Feb 24 are plotted together, windowed in on the fainter, blue region of the spectra where cosmic rays could have the largest effect. \textbf{Right:} Fringing at longer wavelengths, in the same cleaned spectra plotted on the left. Note the significant changes in the interference fringes from one spectral beam to the other.
\label{fig:crclean}}
\end{figure*}

We reject cosmic rays from our raw spectra by first creating a mean spectrum for each object by normalizing all its spectra and then combining them using a trim mean to create a clean master spectrum. We identify bad data in each individual 1-D spectrum using a 5$\sigma$ threshold, where $\sigma$ as a function of wavelength is determined simultaneously with the trim mean.  We correct these bad spectral pixels by replacing them with a scaled value from the normalized mean spectrum for the object in question.  The left-hand panel in Figure \ref{fig:crclean} illustrates the result.

Interference fringing within the thickness of the silicon detector becomes significant at a wavelength of 8200\AA~and continues out to the red limit of GMOS sensitivity. These fringes are not from sky emission, which is cleanly subtracted thanks to the nod-and-shuffle mode of our observations. Instead, the locally monochromatic light from Luhman 16 is interfering with itself in the thickness of the silicon detector, at wavelengths where silicon is no longer sufficiently opaque to absorb all the light on one passage through the detector. This optical interference affects the fraction of light that is actually absorbed by the CCD, and depends in complex ways not only on the local properties (e.g., thickness variations) of the CCD, but also on the spatial illumination pattern of the incoming light. Hence, though fringes with similar morphology appear at the same locations in our spectral flatfield images, the flatfields cannot be used to correct the science data because, due to their more even illumination, the fringes they show have different and typically much lower amplitudes relative to those in the science images. The dependence on local properties of the CCD also means that, at a given wavelength, the fringing pattern is quite different between our two spectral beams (Figure \ref{fig:crclean}, right panel). Because of this, we perform cosmic ray reduction independently on the spectra from each beam. We find that if instead we attempt to clean all the spectra together, high-amplitude fringes are incorrectly rejected as if they were cosmic rays and the spectrophotometry is adversely affected. Note that fringing ceases to be a problem at wavelengths shorter than about 8200\AA, and in particular has no effect in the vicinity of the neutral potassium doublet centered near 7682\AA.

\subsection{Extracting Resolved Spectra}
The very different spectra of Luhman 16 versus the reference stars makes it desirable to obtain relative spectrophotometry by taking the ratio of Luhman 16B's flux to that of Luhman 16A, without using the reference stars at all.  Previous observations have generally indicated that the variations of Luhman 16A, if any, are much smaller in amplitude than those of Luhman 16B: thus, Luhman 16A can serve as a nearby, spectrally matched photometric standard star for studying its highly variable companion, Luhman 16B.

In order to study the variations of Luhman 16B in this way, we must first deconvolve the traces, which are heavily blended especially in our data from Feb 25, when the seeing was less good than on Feb 24. The requirement of high photometric precision necessitates a very careful approach to the deconvolution. In particular, we find that the centroids of the spectral traces must be accurately determined as a function of wavelength (equivalently, x pixel position) across the images. A preliminary fitting of the reference star traces turns out to be essential to deconvolving the blended traces of Luhman 16A and B.

The cross sections of the spectral traces of the reference stars are well-fit by Moffat functions. To perform these fits, we first smooth the spectral traces of the reference stars using a trim mean in a boxcar of width 101 pixels (about 350\AA) in the dispersion (x) coordinate. In each column, we perform a least squares fit over four parameters: y centroid, Moffat $r$, Moffat $\beta$, and spectral flux. As expected, the best-fit Moffat profile is sharpest at the longest wavelengths and becomes blurrier as we move toward the blue, where the seeing is less good. 

Since these four-parameters fits to the reference stars produce good results, we could in principle perform analogous six-parameter fits to the blended traces of Luhman 16A and B: we would simply require y centroids and spectral flux values for two traces rather than just one. In such fits, however, the y centroids of the traces for the two objects are highly unstable, making accurate photometry impossible. Hence, we must forward-model the y centroids of the Luhman 16 traces.

The first step is to precisely measure the separations of the two traces on our images. To do this, we fit the Moffat $r$ and $\beta$ parameters derived for Star 08 (chosen because it is closest to Luhman 16 on the sky) as cubic functions of wavelength.  We adopt these cubic fits as \textit{defining} the Moffat parameters to be used to fit the blended trace of Luhman 16A and B, and fit just the two trace centroids and spectral fluxes. Using a sequence of sharp images from February 24, and constraining the fit to the spectral region where both Luhman 16A and B are bright, we find that the traces are separated by 7.738 pixels (1.131 arcsec), with an RMS variation of only 0.012 pixels (0.002 arcsec).

This accurate value for the separation removes one free parameter from the Luhman 16 fit, since trace centroids for Luhman 16A and Luhman 16B are no longer independent. However, the extracted A/B flux ratio is extremely sensitive to variations in the y coordinate of the trace centroid, and even with the separation fixed, our fits to the centroid are not sufficiently stable as a function of wavelength. Hence, we find it necessary to forward-model the trace centroid as well as fixing the separation. This requires taking into account the effect of differential chromatic refraction (DCR) in Earth's atmosphere, as well as accounting for any optical distortion or misalignment.

To achieve this, we use the traces of all reference stars except the contaminated Star 03. We register them by subtracting the mean y coordinate over the range 7000--7500\AA, and then we average the stellar traces to reduce the noise. The resulting average stellar trace changes from image to image, but in a very predictable way: it can be modeled extremely well by a small, time-invariant linear slant plus the time-dependent contribution from DCR.  Hence, we can model the Luhman 16 trace centroid as a function of wavelength (which we have already mapped to x pixel position), even as the trace curves over time due to DCR.  All that is needed is a precise value for a constant offset in the y coordinate. A slightly different offset must be used for each image, presumably due to guiding errors, gravitational flexure in the telescope, etc.  We find that this constant offset cannot be forward-modeled with sufficient precision from the reference stars, so it is the sole positional parameter that we determine by fitting the Luhman 16 trace directly. We can obtain it with an accuracy of about 0.01 pixels by minimizing the RMS residuals of the double Moffat fit to the Luhman 16 trace. Thus, even though the trace y coordinate for Luhman 16 is a complicated function involving a linear trend added to a continuously changing nonlinear DCR component, we are able to forward-model every part of it except the tiny constant offset.

In our Moffat fits deconvolving the spectra of Luhman 16 A and B, we require $r$ and $\beta$ to be the same for both objects at any given wavelength\footnote{For the final fits, we cannot take Moffat parameters from the reference stars because the latter are too faint at long wavelengths.}. This in itself makes spectrophotometry based on the ratio of the two objects fairly insensitive to the Moffat parameters, but we additionally fit each parameter as a global cubic function of wavelength on each given image --- thus preventing unphysical local `glitches' in the Moffat parameters that might otherwise produce spectrophotometric anomalies.

We note that except for the wavelength calibration, our resolved spectral extraction has almost nothing in common with our unresolved extraction. By design, the unresolved extraction is extremely simple: just a sum over a constant set of rows in the image. As described in this section, the resolved extraction is a sophisticated, customized deconvolution. The large methodological differences between the resolved and unresolved extractions mean that considerably increased confidence may be placed in any results that are confirmed in both.

\subsection{Physically Calibrated Spectra} \label{sec:physcal}

Extracting physically calibrated spectra is not our primary focus, but since we have obtained observations of the A0V star HD 98072 as a telluric standard, we can obtain an approximate physical calibration for our data, with an estimated absolute uncertainty of about 10\%. To do this, we compare our spectrum of HD 98072 with a resampled and blurred model spectrum of Vega (downloaded from \url{http://kurucz.harvard.edu/stars.html}; see, e.g., \citet{Kurucz2011}). We first resample the Kurucz spectrum at 1\AA, and then blur it with a Gaussian of $\sigma = $13\AA, determined empirically to produce the best resolution match to our spectrum of HD 98072. We divide the blurred model Vega spectrum by the observed raw spectrum of HD 98072 to produce a calibration vector that naturally includes both telluric absorption and the spectral response curve of the instrument. As the Kurucz spectrum gives $F_{\lambda}$, and (we assume) the spectrum of Vega is a sufficiently good match to HD 98072, multiplying by our calibration vector converts an observed spectrum into $F_{\lambda}$ times a constant factor related to the ratio of Vega's flux to that from HD 98072. To calculate this constant of proportionality, we use the Vega calibration from \citet{Hayes1985}, which gives a monochromatic flux of 
$3.44 \times 10^{-9}~\mathrm{erg}~\mathrm{cm}^{-2}~\mathrm{sec}^{-1}\mathrm{\AA}^{-1}$ 
for Vega at 5556\AA, and the $V$ magnitude of $9.40 \pm 0.02$ for HD 98072 from \citet{Tycho}. To physically calibrate Luhman 16, we apply an additional scaling to correct from the 5 second exposures of  HD 98072 to the 183 second effective exposure time on Luhman 16.

For our input spectra, we use the average of six HD 98072 spectra taken throughout the night of February 24, and all 26 good science spectra of Luhman 16. Our uncertainty of 10\% is estimated based on the extinction/reddening of HD 98072 (which we have not attempted to correct); possible differences between HD 98072 and Vega; and the imperfect (but fairly close) mean airmass match between our averaged spectra of HD 98072 and Luhman 16. A more precise calibration is beyond the scope of this paper focused on variability analysis. 

\begin{figure*} 
\plottwo{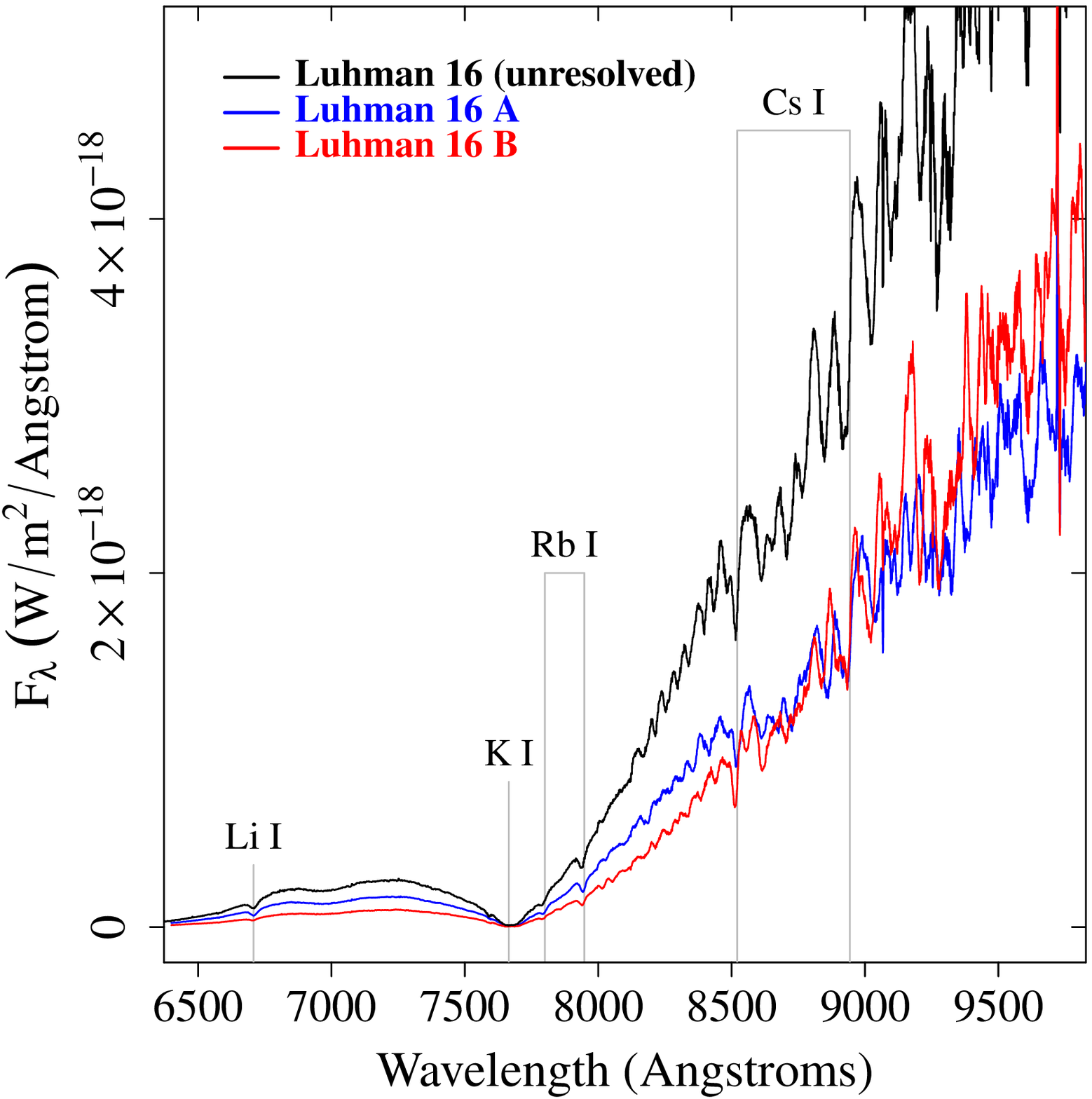}{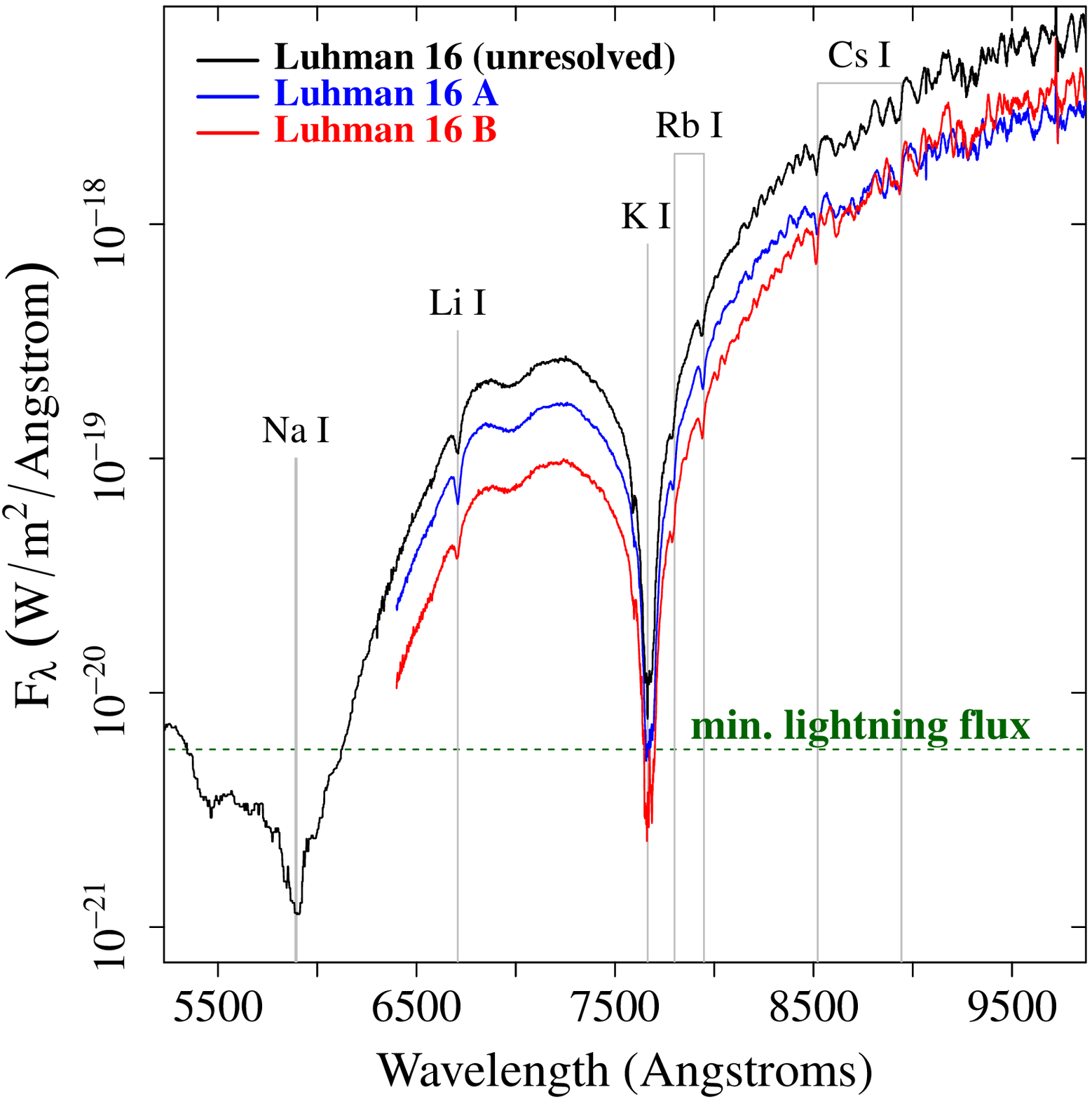}
\caption{Spectra of Luhman 16 from Feb 24, with the individual spectra of A and B deconvolved and physically calibrated as described in the text. {\em Left:} Linearly scaled, showing lines of the neutral alkali metals lithium, potassium, rubidium, and (probably) cesium, though the latter are affected by fringing artifacts. As expected from physical considerations, Luhman 16A is brighter than B at short wavelengths, but at about 9000\AA~they enter the well-known flux-reversal regime (extending to $\sim$ 1.3 $\mu$m) where the T dwarf is brighter despite its lower bolometric luminosity. {\em Right:} Log-scaled, showing the unresolved spectrum at shorter wavelengths where the trace was too faint for deconvolution. The extremely strong sodium D line dominates this regime: hence, all five stable alkali metals have apparently been detected. A lower limit on hypothetical lightning emission (discussed in Section \ref{sec:lightning}) is indicated by the dashed green line.
\label{fig:resspec}}
\end{figure*}

In Figure \ref{fig:resspec} we show the resulting spectra in two forms: first, in a linear scale from 6400--9800\AA; and second, in a log-scale showing the faint blue flux down to 5250\AA. The spectra are dominated, first, by the very red spectral slope of these low-temperature objects, and second by the enormously strong lines of the neutral alkali metals: the sodium D doublet near 5893\AA, and the potassium doublet at 7682\AA\footnote{The sodium doublet consists of two lines at 5890\AA~and 5896\AA, and the potassium doublet of lines at 7665\AA~and 7699\AA, but herein we consistently quote the average wavelength for the two lines in each doublet, because each doublet appears in L and T dwarf spectra as a single, enormously broadened line.}. The H$_2$O absorption band extending redward from 9270\AA~is lost in the interference fringing in Figure \ref{fig:resspec}. This band is known to exist both in telluric absorption and in the intrinsic spectra of L and T dwarfs \citep{Burgasser2008}, and it seems to be detectable in our raw spectra (Figures \ref{fig:rawspec} and \ref{fig:crclean}), but not in the physically calibrated spectra of Figure \ref{fig:resspec}, where telluric absorption has been corrected. The well-known flux reversal between L and T dwarfs is evident in Figure  \ref{fig:resspec}: the T dwarf Luhman 16B is brighter than Luhman 16A at wavelengths longer than 9000\AA. This persists out to about 1.3 $\mu$m (far beyond the coverage of our data), after which the intrinsically more luminous L dwarf again shows a brighter spectrum \citep{Burgasser2013}.

\section{Photometric Validation} \label{sec:val}

We validate our photometry by comparing our GMOS and TRAPPIST results, and our resolved and unresolved GMOS extractions. For these comparisons, we use a wide 8000--10000\AA~bandpass chosen to match approximately the $z'$ filter used for the TRAPPIST photometry. Hence, we sum the spectral flux from our GMOS extractions over this 8000--10000\AA~`red continuum'. We refer to this process of summing extracted spectra over an arbitrary bandpass as `synthetic photometry': the data are real, but we are using spectra to synthesize the photometric measurements that would have been obtained by ordinary imaging photometry with a filter we can customize at will. The remarkable power of spectrophotometry is that we get to choose the `filter set' to optimize scientific return after the data are acquired --- a capability we exploit below in Section \ref{sec:custphot}.

\medskip

We extract synthetic photometry in the 8000--10000\AA~bandpass for the cumulative (unresolved) light from Luhman 16; for each of our reference stars; and for the resolved spectra of Luhman 16 A and B. We find that the ten stars in our MOS masks are photometrically stable with the exception of star 3, which is confused with the negative spectrum of a neighboring star in its sky subtraction region (see Figure \ref{fig:rawdata}). Hence, we use the sum of nine stellar spectra as our reference for unresolved relative photometry of Luhman 16. 

\medskip

Because of the fringing at wavelengths longer than 8200\AA, there is a systematic offset between photometry from beam 1 and that from beam 2 (see Figure \ref{fig:crclean}, right panel). Since we typically switched beams every other image, this systematic offset caused a spurious `zigzag' appearance in our uncorrected photometry. As mentioned above, the interference fringes that cause this are not from the sky background, but from the locally monochromatic light of Luhman 16 interfering with itself inside the CCD chip. The fringing cannot be corrected by flatfielding because it depends in complex ways on the spatial illumination pattern of the incoming light. 

\medskip

We correct the systematic offset between beams by performing a multi-linear fit to the time series data produced by our synthetic photometry. We use a truncated, four-term Fourier series to model the astrophysical variations, and a constant offset to model the systematic difference between beam 1 and beam 2 due to fringing. We perform this fit independently on each night. The fit has 10 free parameters (sin and cos amplitudes for each of four Fourier terms, plus a constant term and the beam offset). Thus, the fit is well constrained by the 26 and 35 input data points that are available, respectively, on Feb 24 and 25. In Section \ref{sec:custphot} below, we will consider the astrophysical meaning of such Fourier fits, but for now the Fourier model's only purpose is to prevent astrophysical variations from biasing our determination of the systematic offset between beam 1 and beam 2, and we discard the model once the offset has been calculated.

The procedure for calculating relative synthetic photometry from our unresolved spectral extractions is therefore:

\begin{itemize}

\item  For each image, add up all the flux in the relevant wavelength bin for Luhman 16 and for each of the nine good reference stars.

\item  Cancel instrumental and atmospheric effects by taking the ratio of the summed flux of Luhman 16 to the sum of all 9 stars.

\item  Normalize the relative photometry constructed in step 2 to have mean value of 1.0 over the time series.

\item  Perform a multi-linear fit to determine the systematic offset between beam 1 and beam 2, using a truncated Fourier series to model astrophysical variations and prevent them from biasing the derived beam offset.

\item Apply the beam offset to the normalized photometry, thereby removing most of the systematic effects from fringing.

\end{itemize}

For synthetic photometry from our resolved spectra, the process is essentially the same (including the correction of beam offsets), but instead of taking the ratio of Luhman 16's total flux to the summed flux of nine reference stars, we take the ratio of Luhman 16 B's flux to that of Luhman 16 A.

\medskip

The beam correction mostly removes the spurious `zigzag' behavior of our photometric time series that was previously imposed by the interference fringing. The fringing is not absolutely constant within each beam (it varies somewhat with seeing and airmass), so residual beam-correlated errors remain at greatly reduced amplitude. We account for these residual errors by including a contribution proportional to the magnitude of the beam offset in our plotted error bars for synthetic photometry.

\begin{figure*} 
\includegraphics[scale=1.1]{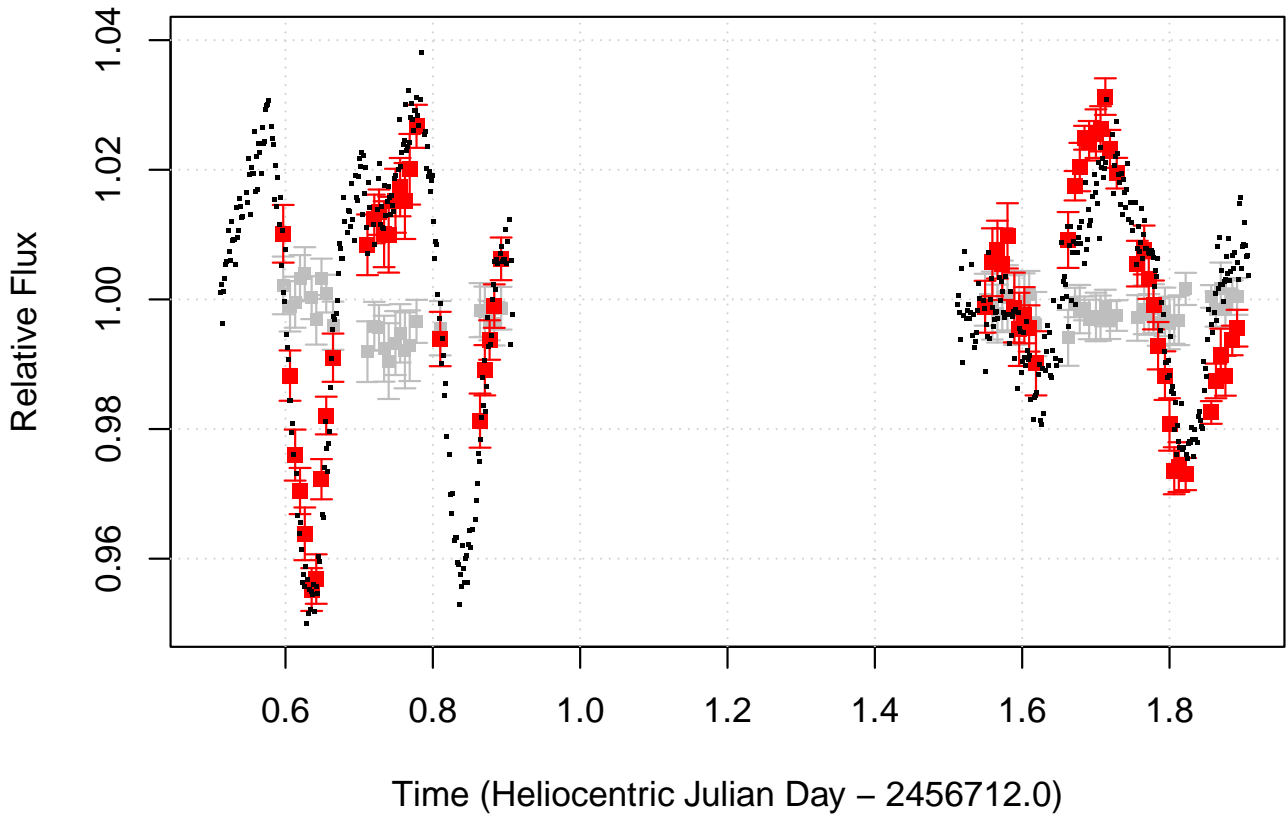}
\caption{Unresolved broadband photometry of Luhman 16 from TRAPPIST (small black squares) and GMOS (large red squares). The GMOS photometry is integrated from 8000-10000 \AA, and is based on the ratio of the Luhman 16 to the summed flux from 9 simultaneously-measured reference stars. For comparison, identically processed photometry from the ratio of one of these comparison stars to the other eight is also shown (gray squares).
\label{fig:TRAPPIST}}
\end{figure*}

Figure \ref{fig:TRAPPIST} compares our unresolved synthetic photometry of Luhman 16 with the simultaneous observations from the TRAPPIST telescope, and with one of our nine GMOS reference stars. The GMOS photometry matches well with the TRAPPIST results, which cover a somewhat larger temporal range. This, together with the nearly constant relative brightness measured by GMOS for the reference star, indicates very low systematic error in our GMOS photometry.

\medskip

\begin{figure*}
\plottwo{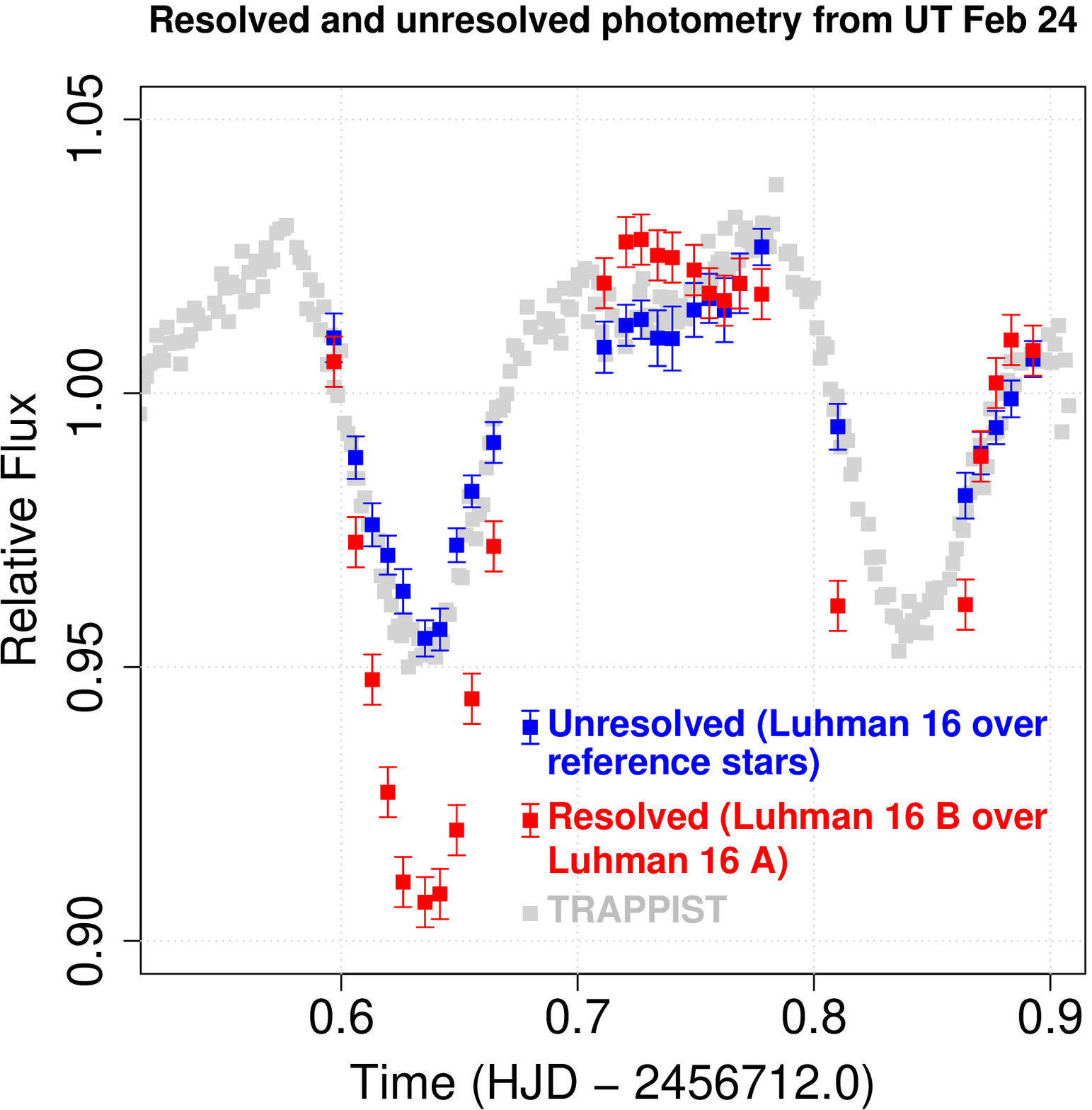}{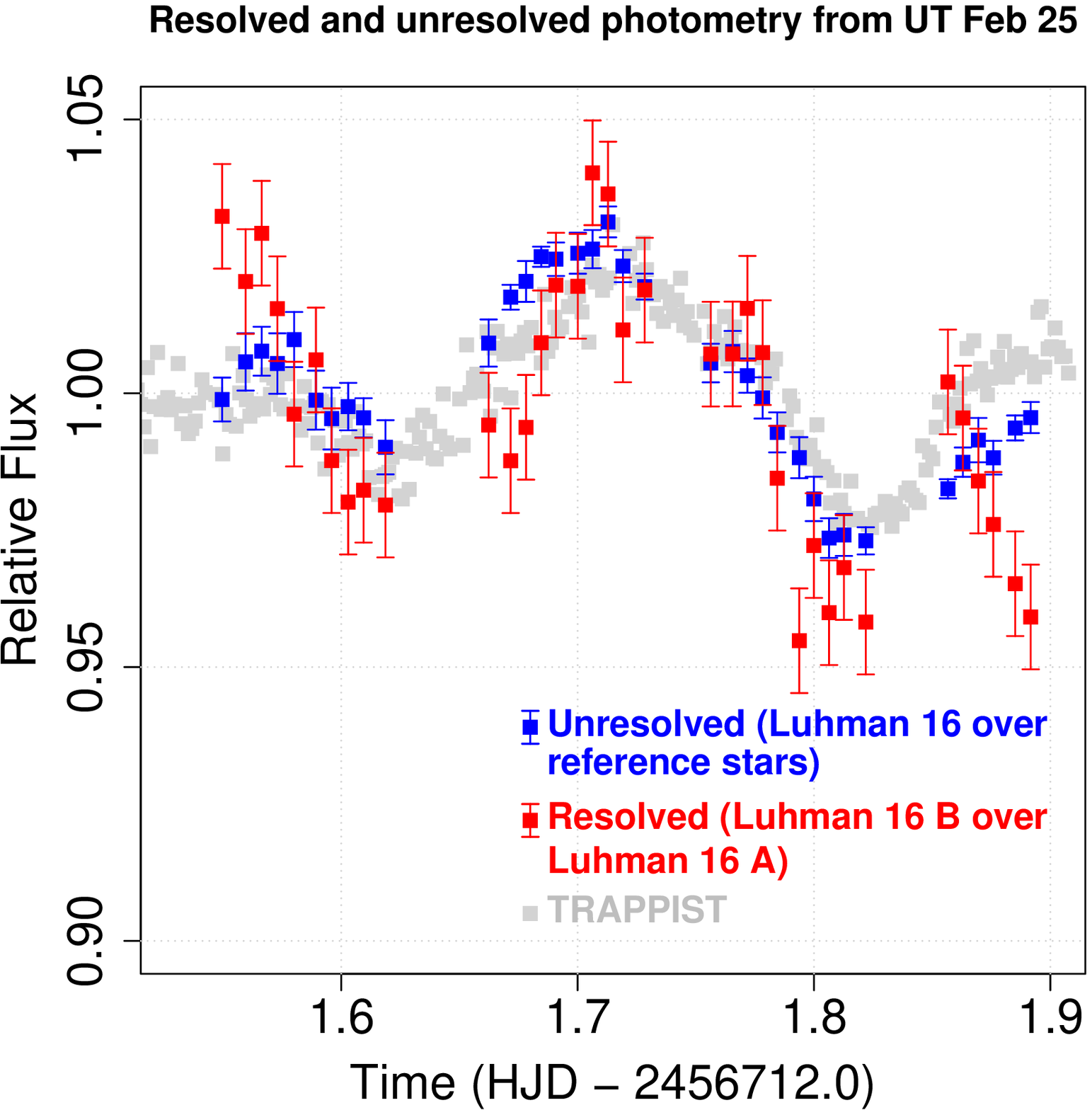}
\caption{Comparison of resolved synthetic photometry of Luhman 16 B with unresolved synthetic photometry of the Luhman 16 system. In both cases the photometry is obtained by integrating over the 8000--10000 \AA~ wavelength range. Under the assumption that Luhman 16B is the dominant source of variability in the unresolved data, the amplitude in the resolved photometry should be twice that in the unresolved data. This is seen in the Feb 24 data, but the situation may be more complicated on Feb 25.
\label{fig:resunres01}}
\end{figure*}

In Figure \ref{fig:resunres01} we compare our resolved and unresolved photometry in the same 8000--10000 \AA~spectral range --- i.e., we compare the normalized flux ratio of Luhman 16 B to Luhman 16 A with the ratio of cumulative light from both to the summed reference stars. In this wavelength range, our measured flux for Luhman 16 B is almost the same as for Luhman 16 A (the mean ratios B/A are 0.996 on Feb 24 and 0.984 on Feb 25). Hence, if Luhman 16 B is the dominant source of variability in the unresolved photometry, the amplitude in the resolved data should be twice as large, while the shape of the light curve should be the same. Figure \ref{fig:resunres01} shows that this expectation is satisfied on Feb 24, but less so on Feb 25, suggesting that Luhman 16 A may have shown some variability on the latter night.

We can probe this further by combining the resolved and unresolved synthetic photometry to solve algebraically for the variations of Luhman 16 A and B. If $u$ is the normalized unresolved data point for a given image, and $r$ is the un-normalized flux ratio $A/B$ for the same image, we solve for the individual relative fluxes of A and B using:

\begin{equation} \label{eq:solveAB}
\begin{array}{lcl}
A + B &=& u
\\ \\
B/A &=& r
\\ \\
A &=& \frac{u}{r+1}
\\ \\
B &=& \frac{ur}{r+1}
\\
\end{array}
\end{equation}

\begin{figure*}
\plottwo{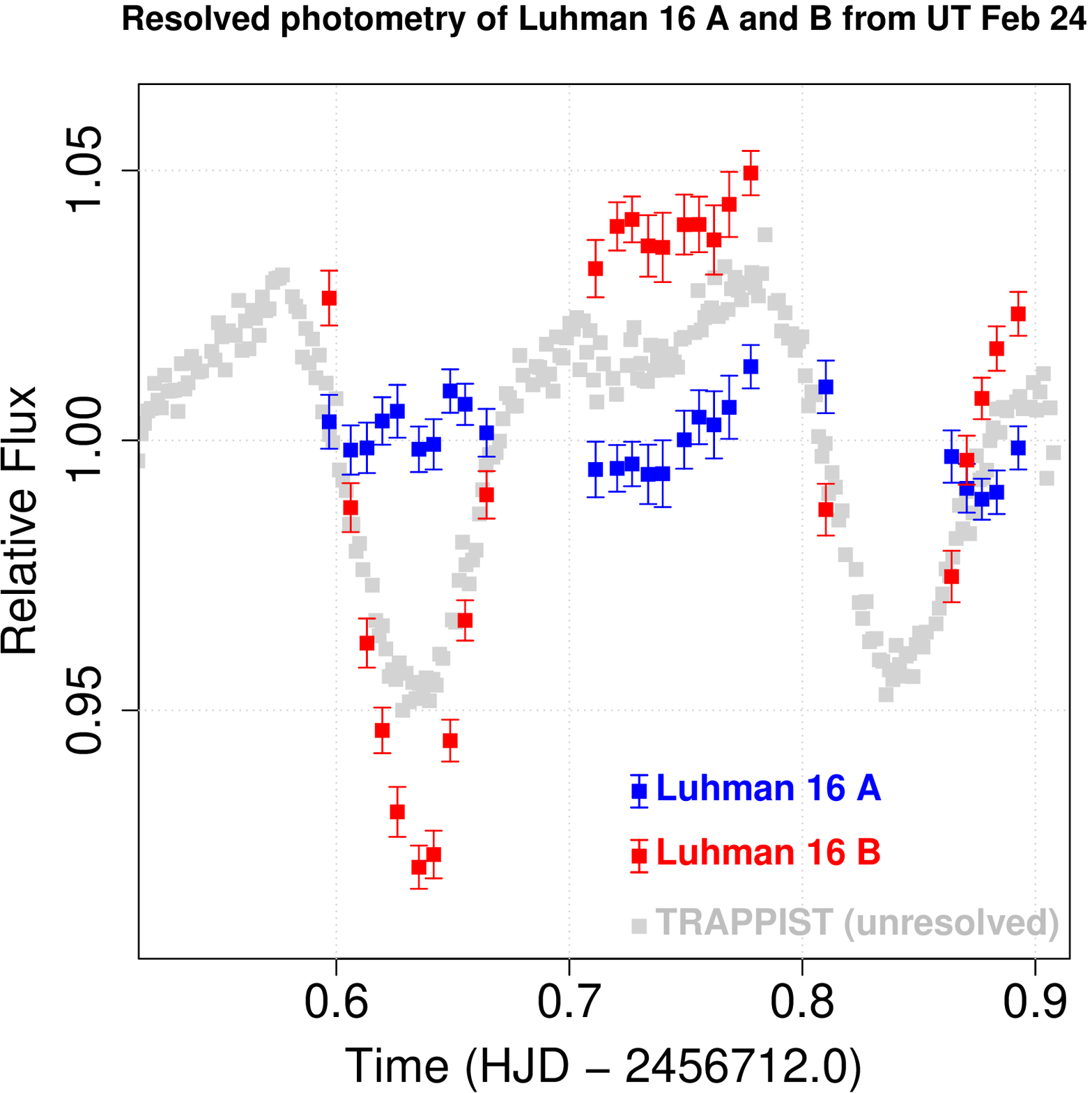}{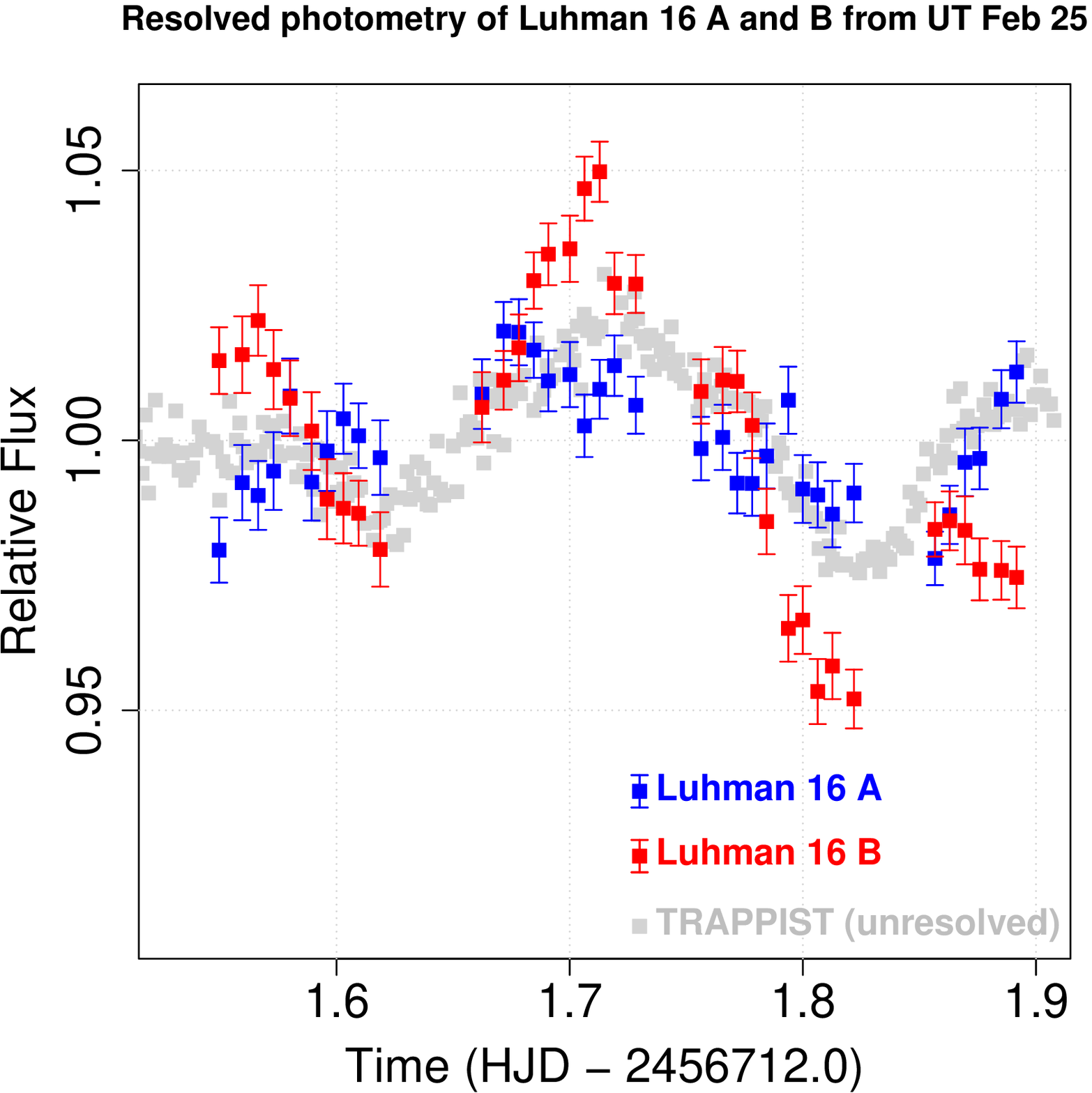}
\caption{Individual variations of Luhman 16 A and B in the 8000--10000 \AA~ wavelength band, calculated algebraically using both resolved and unresolved measurements. On both nights Luhman 16 B has considerably higher amplitude, and is the dominant cause of the variations seen in unresolved photometry as expected. Luhman 16 A is observed to be nearly constant on Feb 24, but significantly variable (albeit less than Luhman 16 B) on Feb 25.
\label{fig:synphotAB}}
\end{figure*}

The results are shown in Figure \ref{fig:synphotAB}. On Feb 24, the variations of Luhman 16 A (if any) are much smaller than those of Luhman 16 B. The amplitude of Luhman 16 B is at least seven times larger. 

As expected based on Figure \ref{fig:resunres01}, the results are less clean on Feb 25. Luhman 16 A does exhibit significant variability, although the amplitude of Luhman 16 B is still about three times higher. The approximation that Luhman 16 A is constant can still be made in dealing with Feb 25 data, but some error will be incurred.

The peak-to-trough amplitude of Luhman 16B in the red continuum is about 13\% on Feb 24 and 7\% on Feb 25.

\section{Wavelength-dependent Specific Amplitudes} \label{sec:specrat}

Having confirmed the basic validity of our spectrophotometry, we search the spectra for wavelength regimes exhibiting particularly interesting behavior. We follow \citet{Apai2013} and \citet{comp} in constructing ratios of spectra representative of the bright and faint states of Luhman 16. Ultimately, we calculate the spectral {\em specific amplitude} (described below), which gives the clearest picture of the wavelength-dependent  photometric amplitude. 

To construct representative bright-state and faint-state spectra, we sort our images from each night based on the relative brightness of the unresolved photometry in the 8000--10000 \AA~red continuum. We then average the spectra in the `bright half' and in the `faint half' of each night. Note that this is slightly different from the procedure of \citet{comp}, where just a few of the brightest and faintest spectra were included in the averages. Our reason for carrying out the average over more spectra in the current case is to reduce the noise in the final result. We create these averaged bright-state and faint-state spectra for both our resolved and unresolved spectral extractions, using a consistent set of `bright' and `faint' images in all cases.

\begin{figure*}
\plottwo{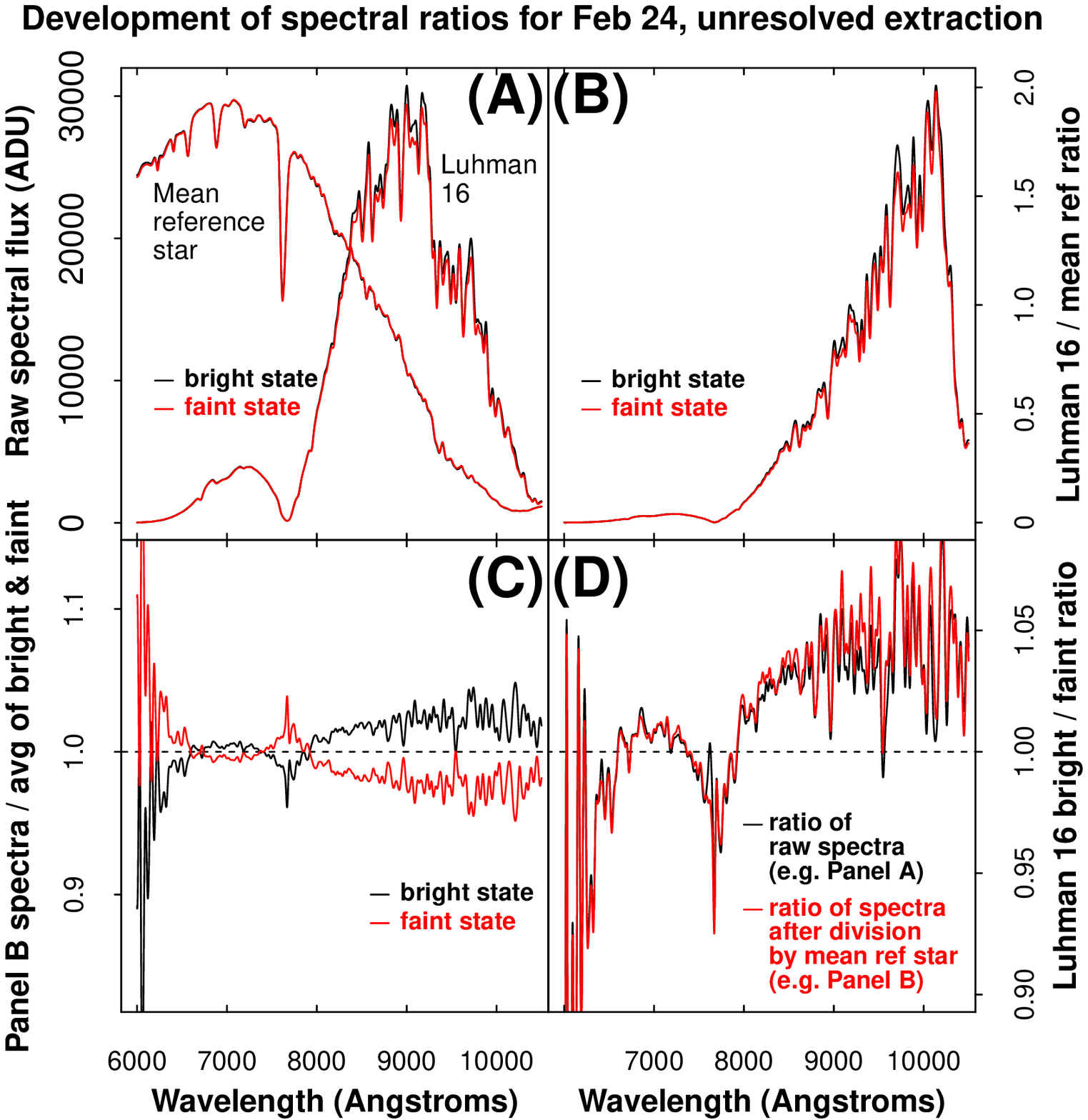}{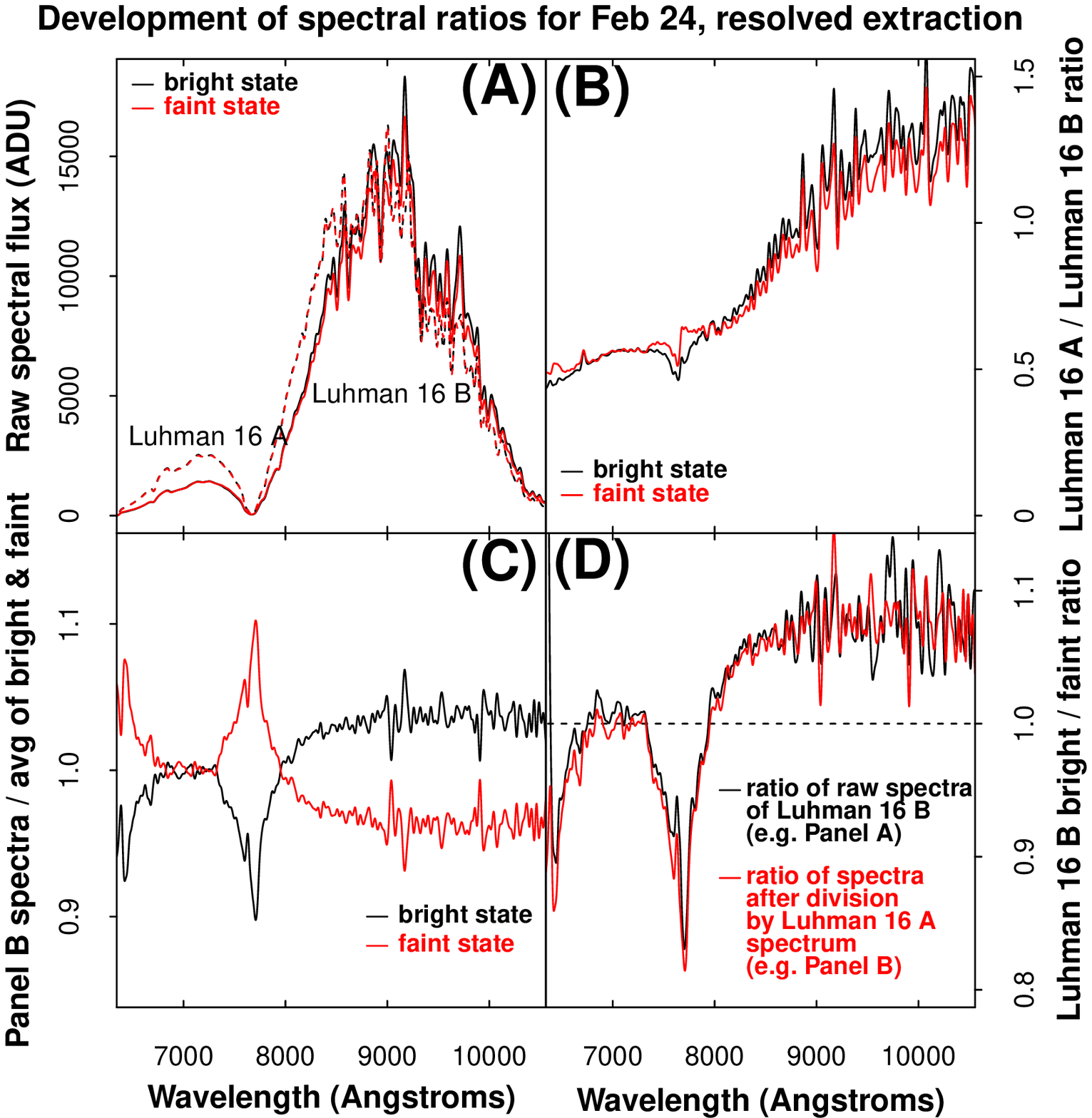}
\caption{As in Figure 4 of \citet{comp}, we show here a sequence of stages leading to our final ratios of Luhman 16 in the bright and faint state, with the unresolved extraction of spectra for the Luhman 16 system (using the average spectrum for our nine reference stars, acquired simultaneously, as a reference) at left and the resolved extraction of the Luhman 16 B spectrum (using Luhman 16 A as a reference) on the right. {\em Panel A:} Raw averaged spectra of target and reference objects in bright and faint states. {\em Panel B:} Bright and faint spectra after removal of telluric effects through division by the reference spectra. {\em Panel C:} Bright and faint spectra from Panel B after division by the average of the two spectra. {\em Panel D:} Bright/faint spectral ratios, without (black) and with (red) removal of telluric effects through division by the reference spectra.\label{fig:specratioexamp01}}
\end{figure*}

Figure \ref{fig:specratioexamp01} illustrates our construction of spectral ratios for the Feb 24 data. We begin with raw spectra of our target (for our unresolved analysis, the combined light of Luhman 16 A and B; for our resolved analysis, Luhman 16 B alone) and a simultaneously acquired reference spectrum (for unresolved analysis, the average of our nine reference stars; for resolved analysis, Luhman 16 A alone). These raw spectra are shown in the (A) panels of Figure \ref{fig:specratioexamp01}. We refer to the bright and faint target spectra as $T_B(\lambda)$ and $T_F(\lambda)$, respectively, and similarly the reference spectra $R_B(\lambda)$ and $R_F(\lambda)$ from the same images. The (B) panels of Figure \ref{fig:specratioexamp01} show the ratios of raw spectra, $T_B(\lambda) / R_B(\lambda)$ and $T_F(\lambda) / R_F(\lambda)$, where the purpose of dividing by the reference spectra is to remove telluric effects. Thus far the differences between the bright and faint spectra are not obvious, but they become so when we construct an average spectrum:

\begin{equation} \label{eq:specrat1}
S_{avg}(\lambda) = 0.5 \frac{T_B(\lambda)}{R_B(\lambda)} + 0.5 \frac{T_F(\lambda)}{R_F(\lambda)}
\end{equation}

\noindent and then divide the individual spectral ratios $T_B(\lambda) / R_B(\lambda)$ and $T_F(\lambda) / R_F(\lambda)$ by this average. This effectively normalizes away the constant features of the spectra, allowing the bright vs. faint differences to appear clearly. The resulting `mean-normalized' bright and faint spectra are shown in the (C) panels of Figure \ref{fig:specratioexamp01}. Finally, we calculate $BF(\lambda)$, the ratio of the bright-state spectrum to the faint-state spectrum, and plot the results in the (D) panels of the figure. If this ratio is formed without first dividing by the reference spectra $R_B(\lambda)$ and $R_F(\lambda)$ (i.e., Equation \ref{eq:BFW}), changes in telluric absorption over the night can cause artifacts. These disappear when the reference spectra are properly applied (Equation \ref{eq:BFR}). Though we do not use the uncorrected form $BF_{uc}(\lambda)$ for any further analysis, we show it (along with the corrected form $BF(\lambda)$) in the (D) panels of Figure \ref{fig:specratioexamp01} to demonstrate that the telluric effects, though detectable, do not dramatically change the result.

\begin{equation} \label{eq:BFW}
BF_{uc}(\lambda) = \frac{T_B(\lambda)}{T_F(\lambda)}
\end{equation}

\begin{equation} \label{eq:BFR}
BF(\lambda) = \frac{T_B(\lambda)/R_B(\lambda)}{T_F(\lambda)/R_F(\lambda)}
\end{equation}

\medskip

\begin{figure*} 
\plottwo{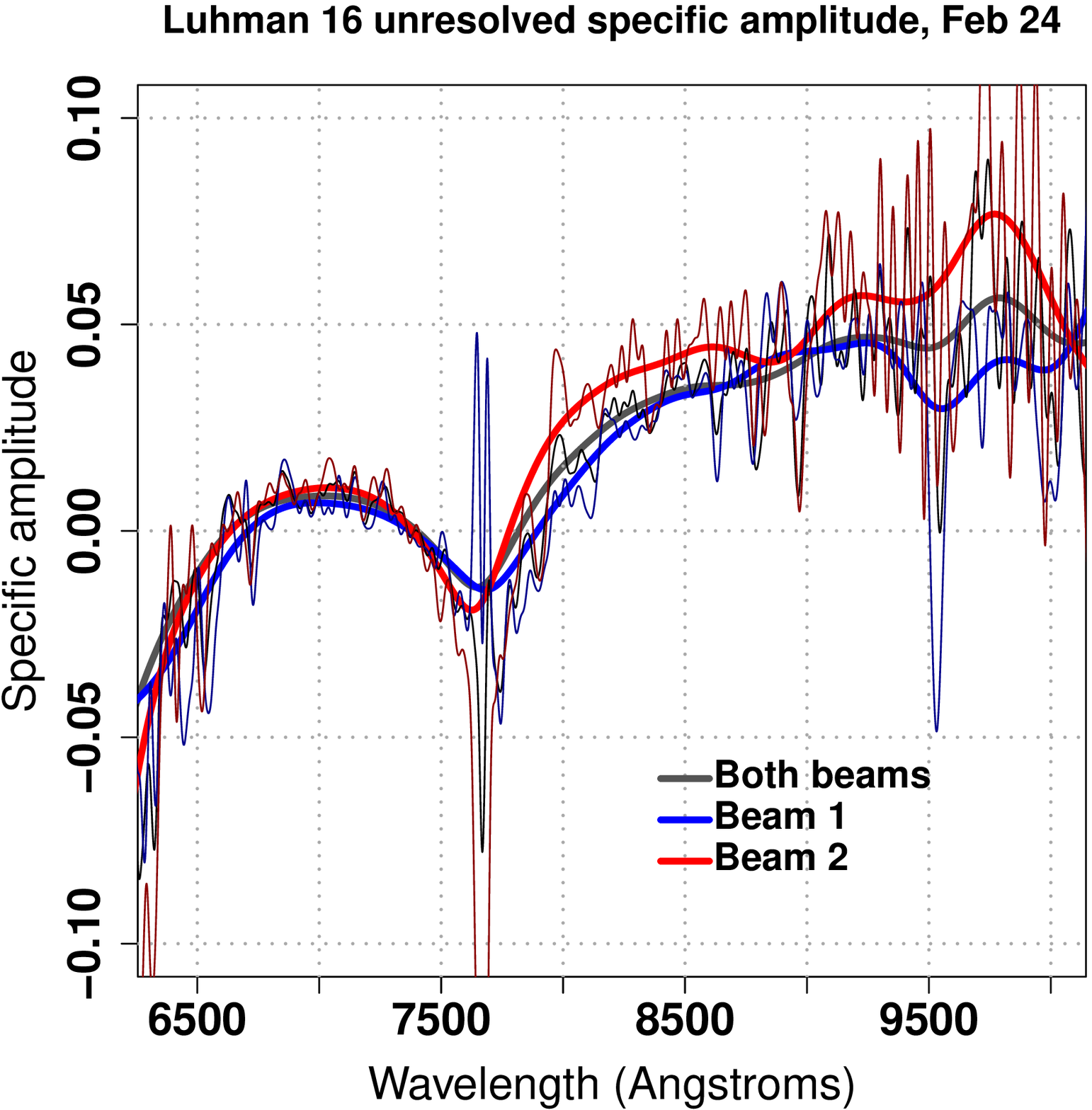}{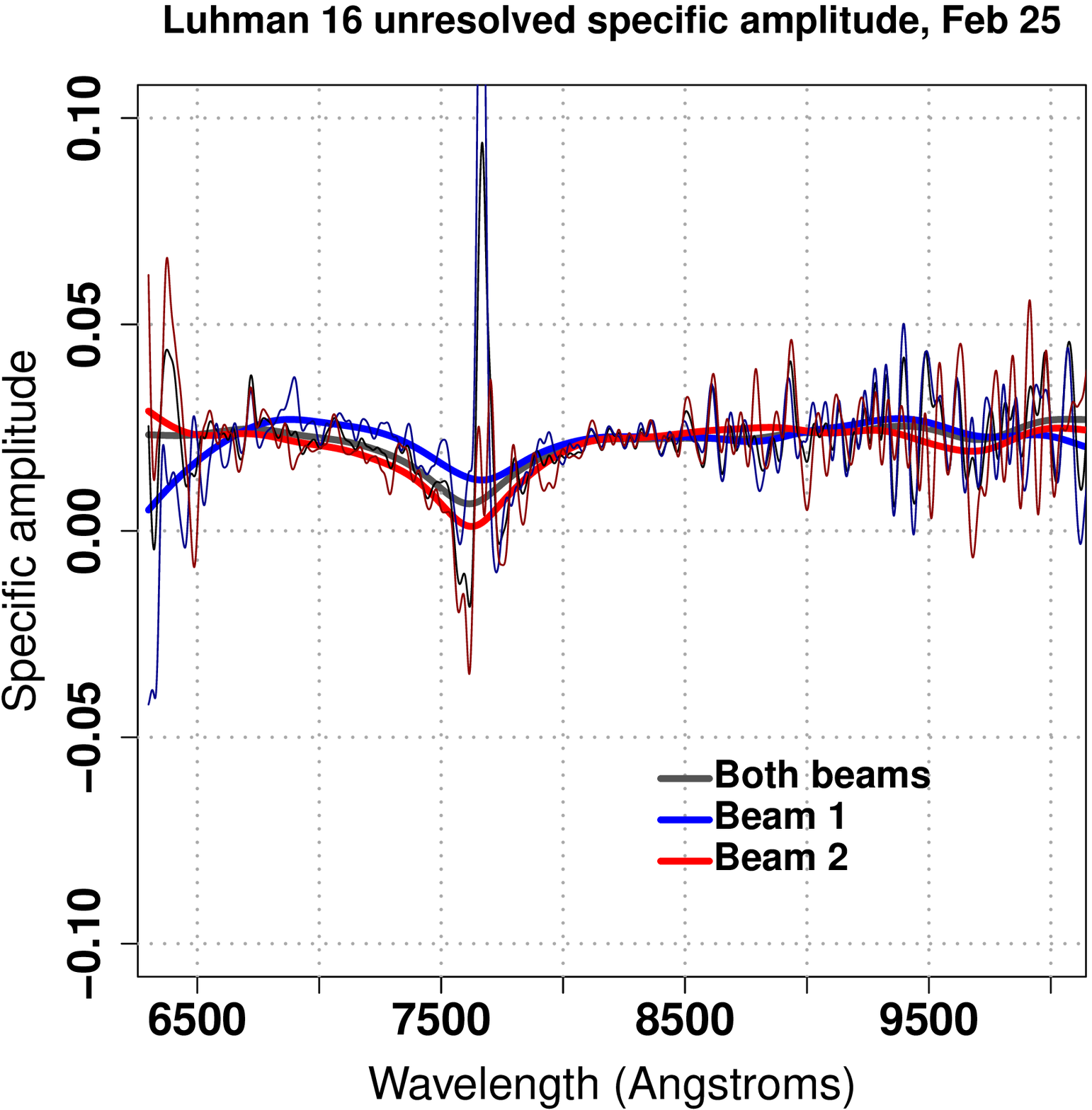}
\caption{Specific amplitudes of Luhman 16's unresolved variability on UT Feb 24 (left) and Feb 25 (right), as calculated using Equation \ref{eq:specrat2}. Thin and thick lines show the data smoothed with kernels of 30\AA~ and 300\AA~ width, respectively. The Feb 24 plot appears noisier because the better seeing on that night produced higher amplitude interference fringes at long wavelengths. 
\label{fig:unres_specamp1}}
\end{figure*}

\begin{figure*}
\plottwo{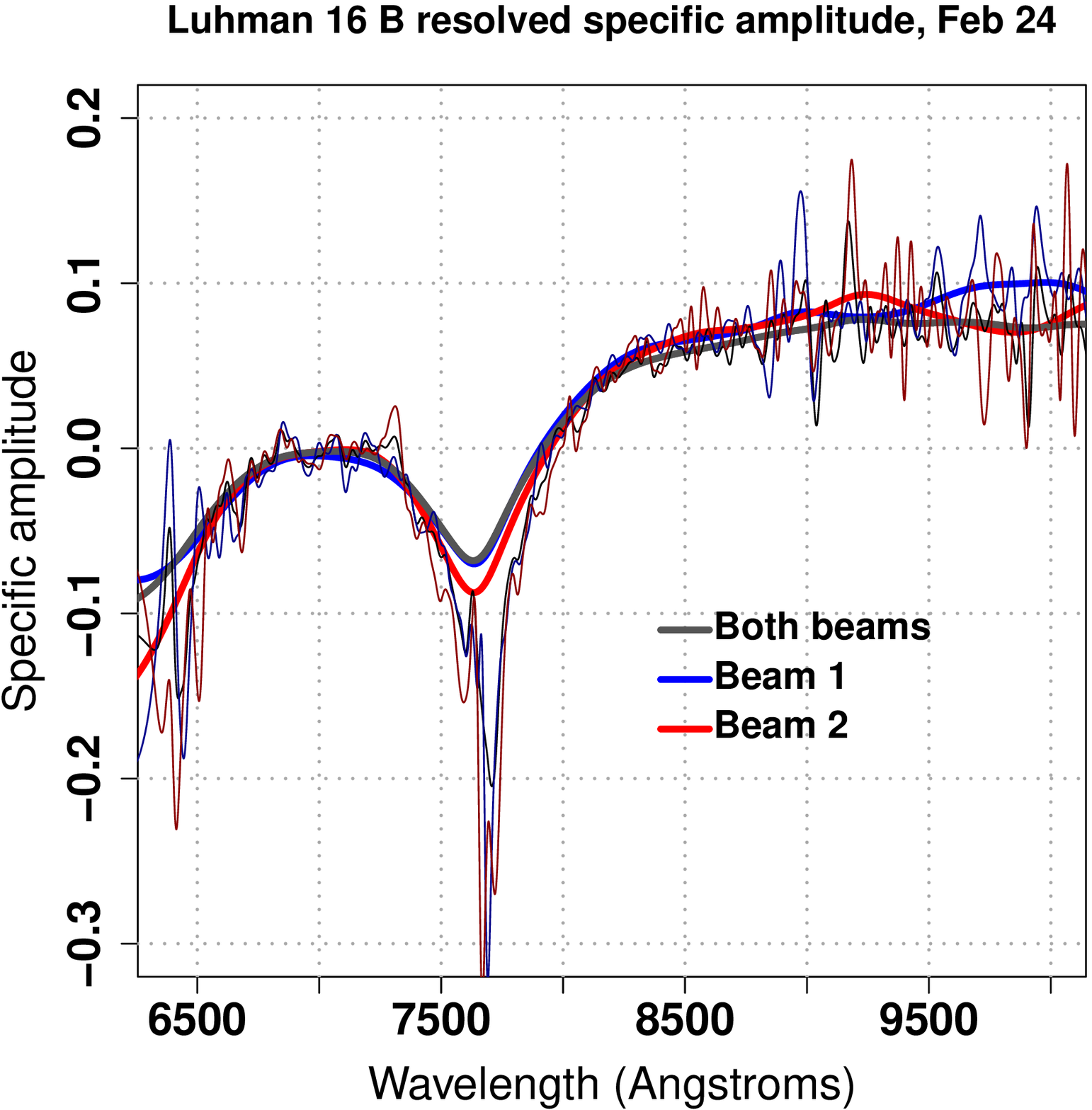}{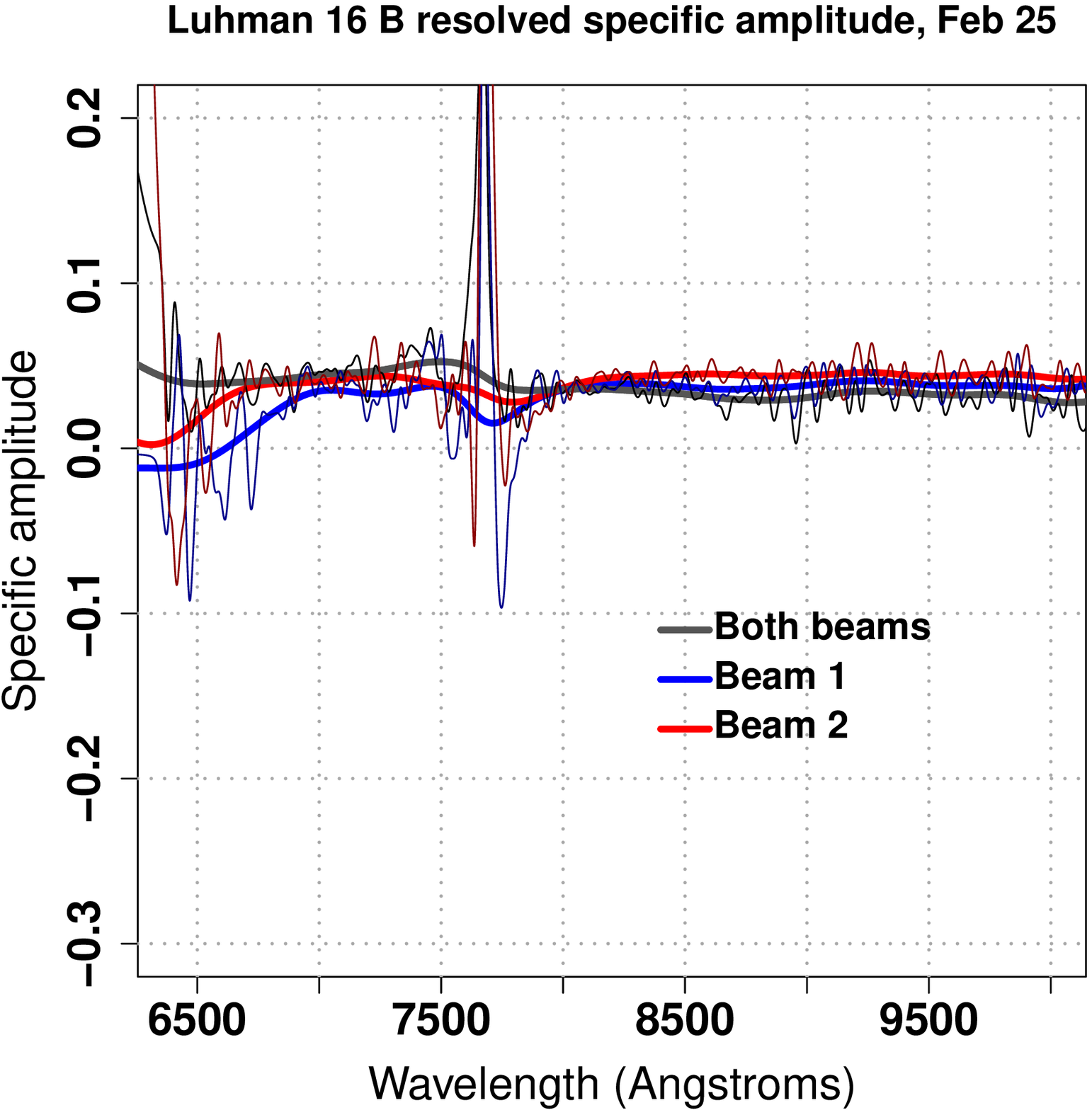}
\caption{Specific amplitude based on resolved spectrophotometry of Luhman 16B on the nights of Feb 24 and Feb 25. Thin and thick lines show the data smoothed with kernels of 30\AA~ and 300\AA~ width, respectively. 
\label{fig:specampres1}}
\end{figure*}

The (D) panels of Figure \ref{fig:specratioexamp01} are exactly analogous to those in the rows labeled `Ratio' in \citet{comp}, Figure 4, and were constructed to enable direct comparison. Herein, we proceed one step further \citep[following Figure 5 in ][]{Buenzli2015a} by defining a quantity we call the {\em specific amplitude}, $A(\lambda)$:

\begin{equation} \label{eq:specrat2}
A(\lambda) = \frac{T_B(\lambda)/R_B(\lambda) - T_F(\lambda)/R_F(\lambda)}{0.5 T_B(\lambda)/R_B(\lambda) + 0.5 T_F(\lambda)/R_F(\lambda)}
\end{equation}

\noindent which is plotted in Figures \ref{fig:unres_specamp1} and \ref{fig:specampres1}. This specific amplitude is simply the wavelength-dependent fractional brightening from the faint state to the bright state: for example, the fact that the resolved specific amplitude of Luhman 16B at 9000\AA~ is about 0.08 means that at this wavelength, the bright-state spectrum of Luhman 16B was 8\% brighter than the faint-state spectrum. Note that since our bright-state and faint-state spectra are each averages of half our images, the specific amplitude is not expected to equal the peak-to-trough amplitude that would be measured photometrically at the same wavelengths. In the case of perfectly sinusoidal variability, $A(\lambda)$ is equal to peak-to-trough amplitude times $2/\pi$, or about 0.63.

\medskip

To confirm the robustness of our specific amplitude results against fringing, we also construct independent bright and faint spectra for each of our spectral beams, and calculate beam-specific versions of $A(\lambda)$ in addition to the master version that combines images from both beams. These are plotted in Figures \ref{fig:unres_specamp1} and \ref{fig:specampres1}. For all three versions, we have smoothed the spectra used to calculate $A(\lambda)$ (Equation \ref{eq:specrat2}) by convolution with a Gaussian kernel with a full width at half maximum (FWHM) equal to 30\AA. This value was chosen as an approximate match to the spectral resolution element, with the objective of smoothing down distracting noise without sacrificing scientifically meaningful spectral resolution. Figures \ref{fig:unres_specamp1} and \ref{fig:specampres1} also show specific amplitude curves calculated with a much larger smoothing kernel of FWHM 300\AA, which are far less noisy but capture the same large-scale behavior.

\section{Strong Anticorrelated Variability in the Alkali lines} \label{sec:anticor}

While the red continuum variability we have observed both with TRAPPIST and GMOS (Section \ref{sec:val}) is entirely unsurprising based on previous results, the wavelength-dependent spectral amplitudes shown in Figures \ref{fig:unres_specamp1} and \ref{fig:specampres1} reveal striking behavior that has not previously been reported. In the Feb 24 data, the specific amplitude becomes strongly negative --- indicating large variations anticorrelated with the red continuum --- over a broad region centered on the 7682\AA~ potassium line. Going toward shorter wavelengths, the specific amplitude returns to zero or small positive values near 7000\AA, before again dropping steeply at the shortest wavelengths we have probed. This behavior is consistent in both the unresolved (Figure \ref{fig:unres_specamp1}) and resolved (Figure \ref{fig:specampres1}) specific amplitude plots, showing little difference except the amplitude dilution in the unresolved analysis due to the nearly constant flux from Luhman 16A.

In strong contrast to these Feb 24 results, our Feb 25 data show much more uniform specific amplitude across the entire wavelength range probed, with relatively small and ambiguous anomalies at the potassium line and the extreme blue.

Comparing the specific amplitude plots for Feb 24 with the physically calibrated spectra of Luhman 16 (e.g., Figure \ref{fig:resspec}) suggests that the departures from the photometric behavior of the red continuum may be caused by the broad absorption lines of the neutral alkali metals (potassium at 7682\AA~and sodium at 5893\AA). As has been previously noted \citep[e.g.,][]{Burrows2002,Burgasser2003}, these lines are extremely strong in L and T dwarfs --- and so broad that their wings span the entire optical spectrum. The degree to which the specific amplitude $A(\lambda)$ at wavelengths outside the red continuum differs from its value in the red continuum correlates well with the strength of the alkali absorption, and this effect is consistently seen (for the Feb 24 data) in Figures \ref{fig:specratioexamp01},\ref{fig:unres_specamp1}, and \ref{fig:specampres1}. By contrast, there is no evidence that the H$_2$O absorption band extending redward from 9270\AA~has any influence on the photometric amplitude on either night --- which is not particularly surprising since this band is not readily visible in our tellurically corrected spectra (Figure \ref{fig:resspec}).

The amplitude of variability in the alkali lines appears to be remarkably large on Feb 24. Figure \ref{fig:specampres1} indicates that the specific amplitude in the potassium line for our resolved Feb 24 data reaches absolute values exceeding the amplitude in the red continuum by more than a factor of two. This means that the variations of Luhman 16B near the core of the potassium line, besides being anticorrelated with those of the red continuum, may have had an amplitude more than twice as great. Given our measurement of a 13\% red continuum amplitude from Section \ref{sec:val}, this suggests the amplitude could have been 30\% or even more in the core of the potassium line (note that, because of our averaging over faint-half and bright-half spectra, the specific amplitude plotted in Figure \ref{fig:specampres1} is always less than the full peak-to-trough amplitude at a given wavelength). To explore this further, we employ the synthetic photometry methods presented in Section \ref{sec:val} in a set of bandpasses informed by our specific amplitude plots.

\section{Probing Alkali Absorption with Synthetic Photometry} \label{sec:custphot}

While in Section \ref{sec:val} we validated our synthetic photometry using a wide spectral bandpass aimed at matching the TRAPPIST data, we now opt for narrower bands to probe the wavelength dependence of Luhman 16's photometric variability. This approach complements our specific amplitude analysis by revealing the time-resolved photometric behavior instead of merely probing the difference between the average bright-state and faint-state spectra.  We select six wavelength bands (Table \ref{tab:bands}), chosen to span our useful spectral coverage while also specifically targeting the alkali lines.

\begin{deluxetable}{llll}
\tablewidth{0pt}
\tablecaption{Synthetic Photometry Bands\label{tab:bands}}
\tablehead{\colhead{Wavelength range} & \colhead{Description} & \colhead{Alkali effects}}
\startdata
6000--6400\AA & Sodium red wing & Very strong\\
6400--7400\AA & Blue hump & Strong\\
7530--7830\AA & Potassium line & Very strong \\
7800--8500\AA & Potassium red wing & Moderate \\
8500--9270\AA & Red continuum I & Weak \\
9270--9800\AA & Red continuum II & Weak \\
\enddata
\end{deluxetable}

\begin{figure*}
\plottwo{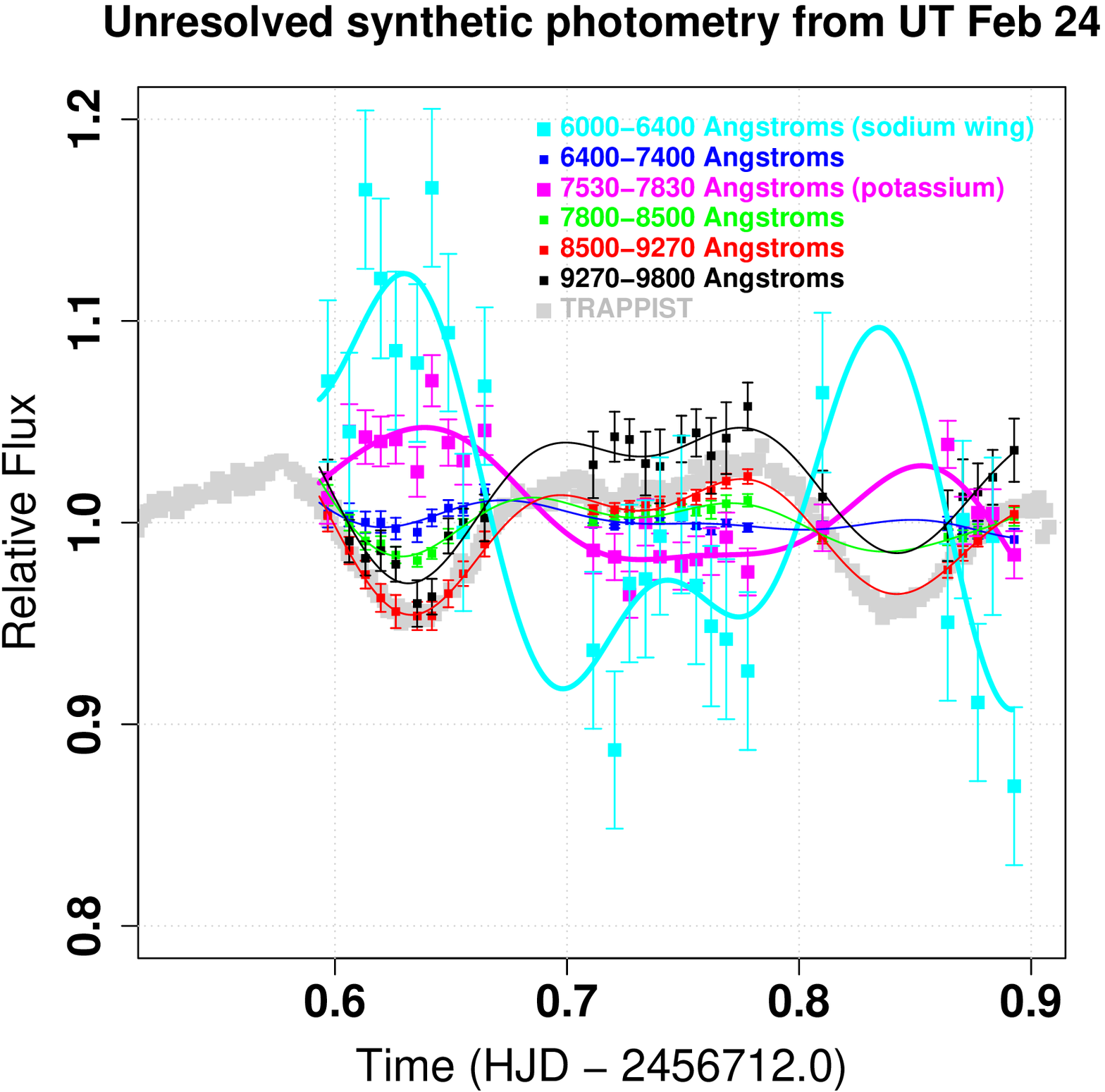}{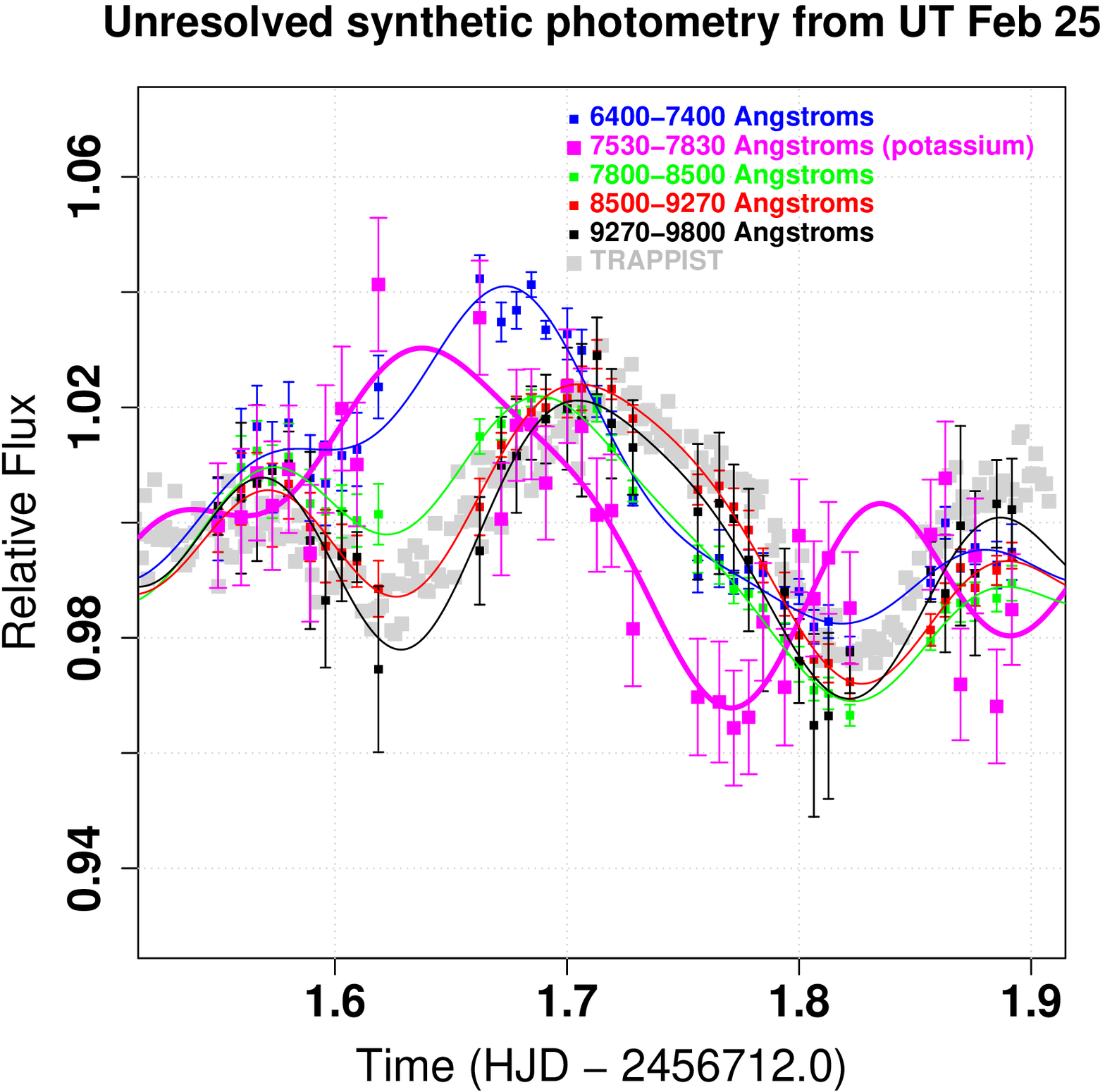}
\caption{Synthetic photometry of Luhman 16, unresolved (i.e., cumulative flux from both Luhman 16 A and B) in bands chosen to span the wavelength range of our GMOS data while emphasizing regions that appeared interesting in our analysis of spectral specific amplitudes. {\em Left:} Our Feb 24 data, including the 6000--6400 \AA~range where Luhman 16 becomes very faint and resolved measurements were impossible. {\em Right:} Data from Feb 25, excluding the shortest-wavelength range, which was rejected by the order-blocking filter we used on that night. Spectral regions affected by the strong, hugely broadened lines of potassium (7682\AA) and sodium (5893\AA) show photometric behavior that deviates from that of the red continuum (8000--10000 \AA) to an extent that correlates with the strength of the alkali absorption.
\label{fig:synunres}}
\end{figure*}

\begin{figure*}
\plottwo{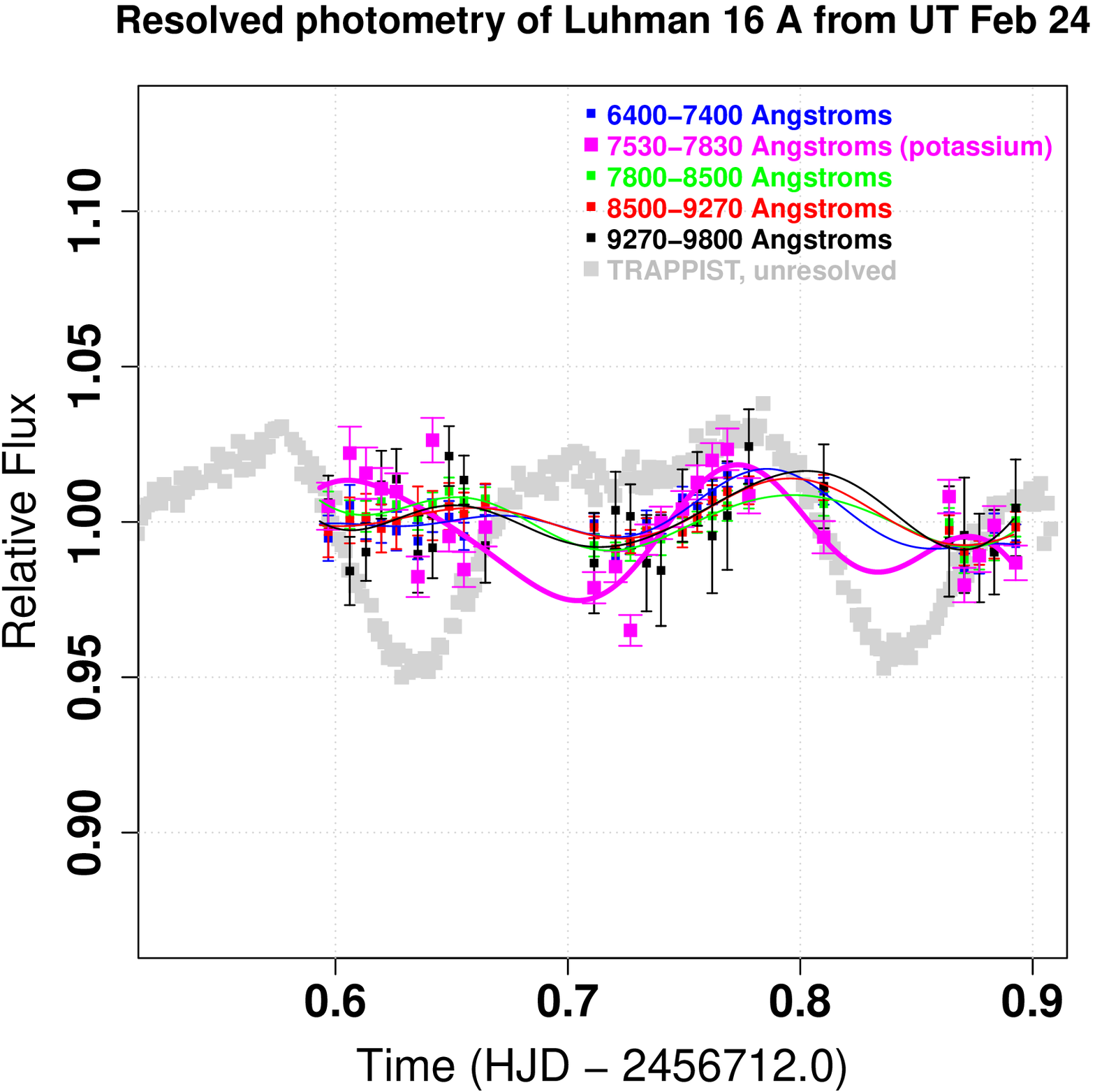}{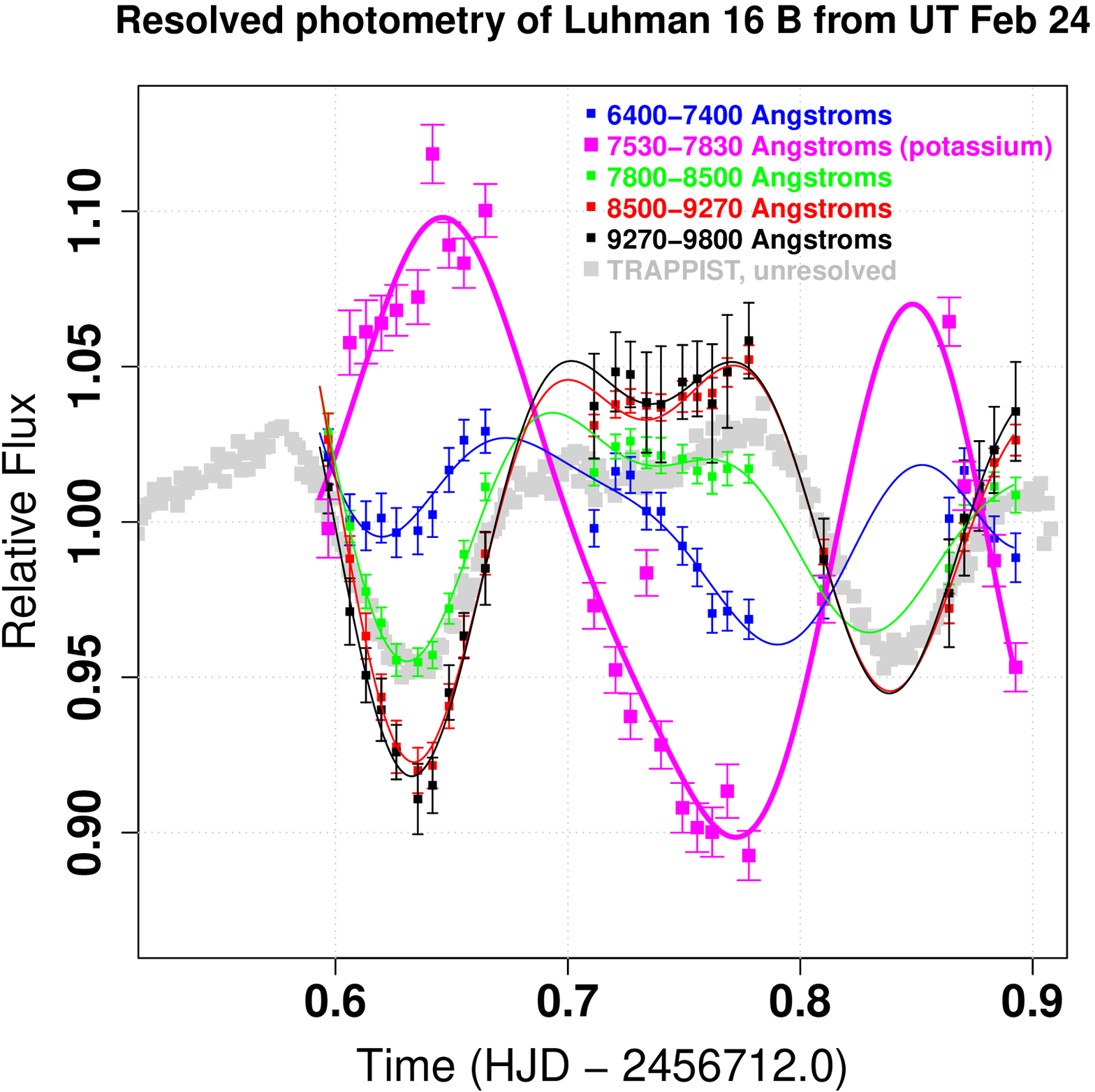}
\caption{Individual flux variations of Luhman 16 A and B on Feb 24 in five different wavelength bands, matching those of Figure \ref{fig:synunres} except for the 6000--6400 \AA~band where the spectra were too faint to be resolved by deconvolution. Luhman 16 B is much more variable than Luhman 16 A, and although some differences exist between the photometric variations seen here in Luhman 16 B and those shown in Figure \ref{fig:synunres} for the total system light, these resolved measurements corroborate the unresolved detection of anomalous photometric behavior correlated with the strength of the alkali absorption.
\label{fig:synAB1}}
\end{figure*}

\begin{figure*}
\plottwo{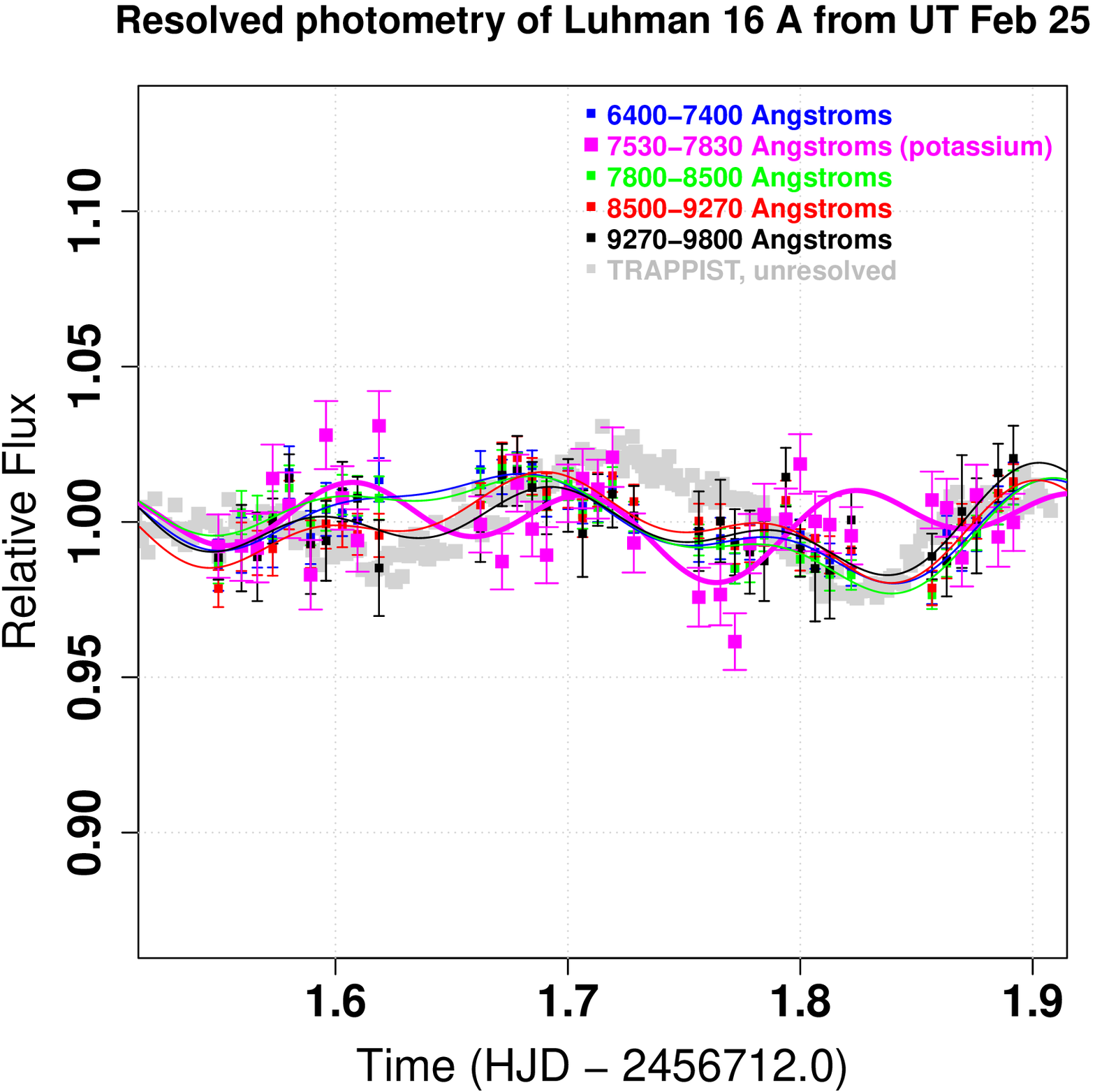}{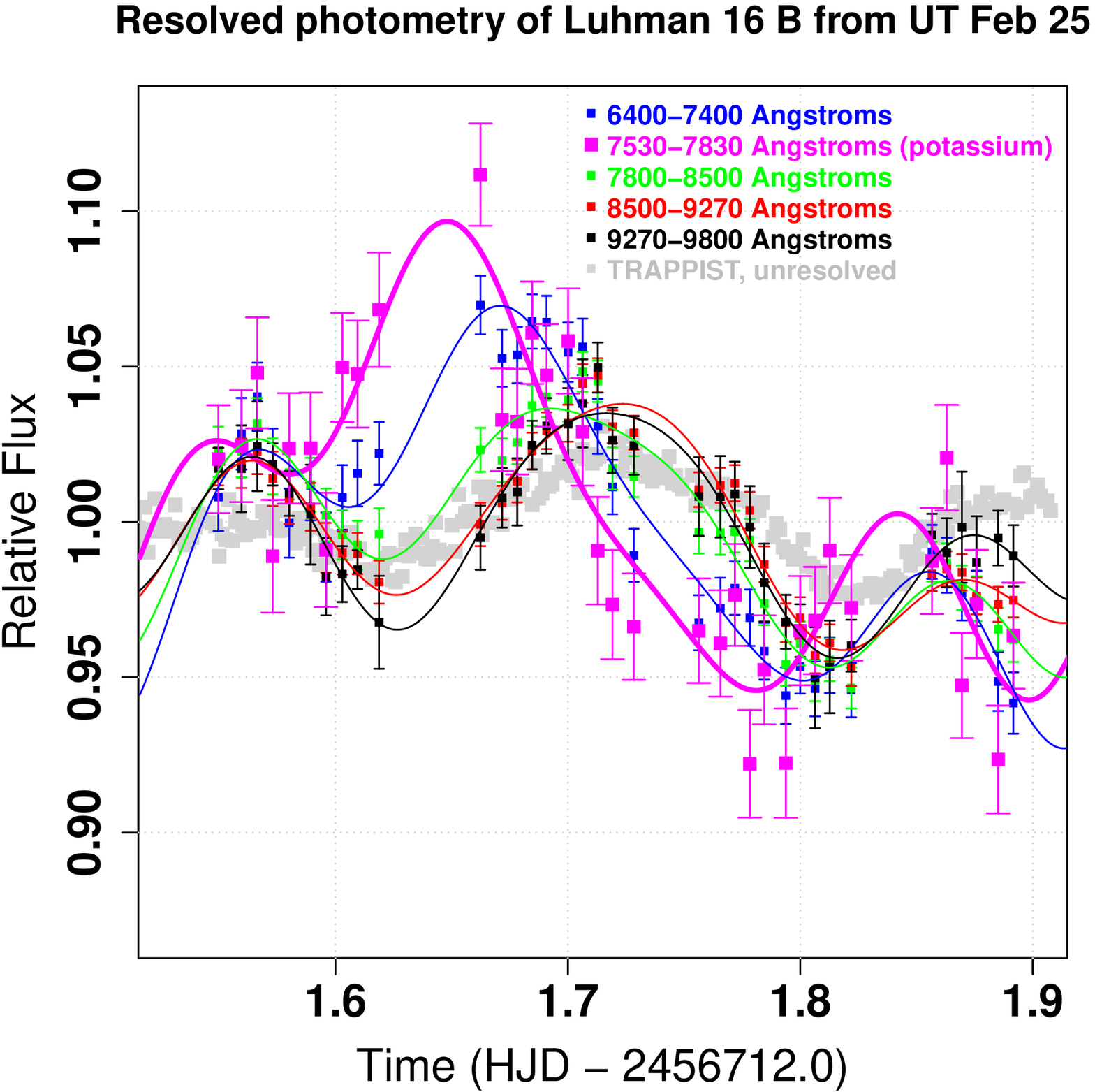}
\caption{Individual flux variations of Luhman 16 A and B on Feb 25 in five different wavelength bands, matching those of Figure \ref{fig:synunres}. While Luhman 16 A is not as photometrically constant on this night, the resolved observations nevertheless corroborate the unresolved detection of anomalous photometric behavior correlated with the strength of the alkali absorption, and show that the anomaly comes primarily from Luhman 16B as expected.
\label{fig:synAB2}}
\end{figure*}

Four of our bands need more description than Table \ref{tab:bands} provides. The longest one, 9270--9800\AA, probes the part of the red continuum that could be affected by H$_2$O absorption; however we do not see evidence of such effects. The 7530--7830\AA~ band is centered on the potassium doublet and extends just far enough into the wings to get a reasonable amount of flux. The 6400--7400\AA~band spans the `blue hump' between the enormous troughs of the potassium doublet and the 5893\AA~sodium doublet --- but remains strongly affected by alkali absorption since the wings of the two lines overlap \citep{Burrows2000,Burrows2002}. Our shortest band, 6000-6400\AA, which extends deep into the trough of the 5893\AA~sodium doublet\footnote{We cannot probe a band {\em centered} on the 5893\AA~sodium line as we did for potassium at 7682\AA, because Luhman 16 is too faint in the sodium line.}, can be probed only in the unresolved Feb 24 extraction. It cannot be analyzed in resolved data because the flux from Luhman 16 was too faint for deconvolution; nor in the Feb 25 data because some of the flux was excluded by the order-blocking filter. In Figure \ref{fig:synunres}, we show our unresolved synthetic photometry for both nights, while Figures \ref{fig:synAB1} and \ref{fig:synAB2} show resolved photometry for Luhman 16 A and B, extracted algebraically using Equation \ref{eq:solveAB}, for Feb 24 and Feb 25 respectively.

We note that the presence of the telluric oxygen A absorption line at $\sim$7600\AA~could affect our unresolved synthetic photometry in the 7530--7830\AA~potassium band. Taking the ratio of Luhman 16's flux in this band to that of the reference stars might not eliminate all telluric effects, because the flux from Luhman 16 is distributed so differently across this band relative to that from the stars that the fractional effect of the oxygen A absorption might be substantially different. This effect can be removed by dividing Luhman 16 by the reference star spectrum before summing over the photometric band. We have performed this experiment, and we find that although the resulting photometry is not identical, the differences are insignificant and could not affect any of our scientific interpretation. Since dividing by the reference spectrum before summing over the bands increases the random noise --- and is in any case a rather strange hybrid between synthetic photometry and the spectral ratios of Section \ref{sec:specrat} --- we have used only normal synthetic photometry for Figures \ref{fig:synunres} through \ref{fig:synAB2}, and for our final scientific interpretation.

In each of these synthetic photometry plots, we also show a four-term Fourier fit based on a 10-hour master period (that is, a linear least-square fit including sine and cosine terms with periods of 10.0, 5.0, 3.33, and 2.5 hr). The master period was chosen to be about twice the actual rotation period of Luhman 16B, but we do not argue that the sub-periods have astrophysical meaning --- rather, the truncated Fourier model is simply a functional form able to capture most of the astrophysical variations without overfitting. Since the 10-hour master period is longer than our monitoring interval on each night, the Fourier model could in fact capture {\em arbitrary} photometric variations over this period, as long as they contained no power at frequencies higher than that corresponding to the shortest, 2.5 hr period in the model. Hence, the model should be able to fit the weak variability of Luhman 16A, even though its period is believed to be 8 hours rather than 5 \citep{Mancini2015,Apai2021}.

The data from Feb 24, both resolved and unresolved, are nicely consistent with our idea that the degree to which the photometric behavior at a given wavelength differs from that in the red continuum is determined by how strong the alkali absorption is at that wavelength. Of our six wavelength bands, the two longest (8500--9270\AA~and 9270--9800\AA) vary in lockstep with the red continuum, the H$_2$O absorption in the longer band having no apparent photometric effect. The 7800--8500\AA~band shows variations correlated with the red continuum but at reduced amplitude; the 6400--7400\AA~band has even smaller amplitude; the 7530--7830\AA~potassium band shows large-amplitude variations anticorrelated with the red continuum, and the extreme blue (deep in the sodium line) shows even larger anticorrelated variations. The resolved photometry of Luhman 16B in the potassium band shows variations with an amplitude of 20\%. Since the band extends into the wings of the potassium line in order to get sufficient flux, this 20\% amplitude measurement is entirely consistent with our suggestion in Section \ref{sec:specrat} that the amplitude could be 30\% or even more in the core of the line.

The Feb 24 synthetic photometry results might have been anticipated from the specific amplitude plots in Section \ref{sec:specrat}. What is more interesting is that the synthetic photometry from Feb 25 {\em also} shows photometric behavior that differs from that of the red continuum in proportion to the strength of the alkali absorption. This could not have been predicted based on the specific amplitudes, because on Feb 25 the alkali effect does not consist primarily of an amplitude reversal -- rather, it appears more like a phase shift. Hence, the two longest wavelength bands vary together; the 7800--8500\AA~band appears to be shifted slightly earlier in time; the 6400--7400\AA~`blue hump' shifts further, and the potassium band further still.  Hence, on both nights there is evidence for an alkali effect --- but it is not the same effect.

Our Feb 24 observations are not the first evidence of a photometric amplitude inversion for Luhman 16B in the optical potassium and sodium lines. \citet{Biller2013} monitored Luhman 16 in the near-infrared $J$ and $H$ bands and the optical $r'$, $i'$, and $z'$ bands. In the $J$ and $H$ bands, as well as in $z'$ (which lies squarely on our red continuum) they found correlated variations at similar amplitudes. However, in the $r'$ and $i'$ bands, which include the the 5893\AA~sodium doublet and the 7682\AA~potassium doublet respectively, they found highly significant variability anticorrelated with $z'$, $J$, and $H$. In light of our spectrally resolved results, it seems clear that \citet{Biller2013} were in fact the first to detect the inversion of Luhman 16B's photometric amplitude in the optical absorption lines of the neutral alkali metals, although without spectra they could not recognize this.

The $r'$ and $i'$ band photometric amplitudes measured by \citet{Biller2013} are similar to or less than their $z'$, $J$, and $H$ amplitudes, consistent with the dilution of the strong alkali signatures by the relatively broad $r'$ and $i'$ bands. In contrast to our own observations of an amplitude inversion on Feb 24 but not Feb 25, \citet{Biller2013} observed it on two different nights. This suggests that even though our Feb 25 data show the atmosphere of Luhman 16B is not always in a state that will produce an amplitude inversion in the optical alkali lines, it is not rare to find it in such a state. Similarly, \citet{Buenzli2015b} detected a large reduction in the amplitude of Luhman 16B in the red wing of the 7682\AA~potassium doublet, though their data did not extend to short enough wavelengths to see if there was an amplitude inversion in the core of the line.

\section{Physical Interpretation: A Simple Conceptual Model} \label{sec:expectation}

Rotationally modulated variability such as that of Luhman 16 requires some type of longitudinal inhomogeneity. Since clouds are known to be important in brown dwarf atmospheres, the inhomogeneous distribution of them has long been considered the likeliest explanation for brown dwarf variability. A simplified but useful starting point for considering such variability is to imagine that the atmosphere is homogeneous except for the clouds --- or more accurately, that whatever longitudinal variations may exist in other atmospheric properties such as the temperature profile and gas composition, it's mainly the clouds that produce the rotational variations. 

This conceptual picture makes the qualitative prediction, already mentioned in Section \ref{sec:photintro}, that the amplitude of variability should be low in spectral regions where the gas opacity is high. At such wavelengths, the opacity prevents us from seeing deep into the atmosphere even in the absence of clouds, so cloudy and non-cloudy regions appear uniformly faint, producing little rotational variability. On the other hand, at wavelengths with low gas opacity --- and in atmospheric regions with little cloud --- we can see into deep, warm layers of the atmosphere. Such regions should be very bright, while areas with thick clouds extending to high altitudes will be much darker, and the strong contrast should produce large-amplitude rotational variability at these wavelengths. The expectation, then, is of high photospheric contrast and high-amplitude variability in spectral regimes with low gas opacity, and low contrast and low-amplitude variability where the gas opacity is high (i.e., in the strong spectral lines). Although this picture is simplistic (e.g., real clouds would alter the vertical temperature profile of the atmosphere, and possibly the chemical abundances and gas opacity also), its basic prediction has been verified in many cases -- e.g., \citet{Apai2013}, who found that the variability amplitude of 2M2139 and SIMP0136 was greatly reduced in the region of strong H$_2$O absorption from 1.35--1.50$\mu$m; \citet{Buenzli2015a}, who found the same thing for Luhman 16B; and the finding of \citet{comp} that the variability amplitude of Luhman 16B on Feb 24 was lower in this same H$_2$O absorption band.

Our optical data from Luhman 16 on Feb 24 do {\em not} fit into this simple conceptual picture, because they show not merely an amplitude decrease but an amplitude \textit{reversal} in spectral regions affected by strong alkali absorption. When regions of thick cloud rotate into view, the flux in low-opacity spectral regions (e.g., the red continuum) would be expected to get much fainter, while the flux in high-opacity regions (e.g., the alkali lines) should stay the same or get only a little fainter. What we actually see is that when the red continuum gets faint (presumably because clouds are rotating into view), the flux in the alkali lines {\em brightens}. In retrospect, it seems clear that \citet{Biller2013} also observed the same phenomenon in broadband photometry on the nights of UT 2013 April 16 and 22.

\citet{comp} saw further deviations from the simplified model's predictions: on the same night (Feb 24) that we detected the amplitude inversion in the optical alkali lines, they made a highly significant detection of an {\em increase} in the amplitude of variation in the potassium doublets near 1.173 $\mu$m and 1.248 $\mu$m. Once again, the simple expectation is that at wavelengths where gas opacity is high (i.e., spectral lines), the photometric amplitude will be lower. For Luhman 16B on February 24, this expectation is defeated in different ways for different lines of the same element: in the infrared potassium lines, \citet{comp} reported not a decrease in photometric amplitude but an increase, while in the optical potassium doublet we see not merely a decrease but an inversion. 

Clearly, the simplified picture does not describe the variability of Luhman 16B in the alkali lines. Does the variability result from inhomogeneous cloud obscuration at all? Suggestions have been made of auroral or lightning emission in brown dwarfs \citep[e.g.,][]{Helling11a,Hallinan07,Berger09}. Longitudinal inhomogeneity in such sources would cause rotational variability, at least at some wavelengths. We discuss such interpretations of our data below in Section \ref{sec:emm}, but ultimately conclude that these non-photospheric phenomena likely could not produce the behavior we have observed, and that more sophisticated cloud based models (Section \ref{sec:interp}) are the most probable explanation.

\section{Ruling Out Non-Photospheric Emission} \label{sec:emm}

Suppose the atmosphere of Luhman 16B contains a source of flux with far less luminosity than the photosphere but a very different spectrum --- specifically, a spectrum without deep alkali lines, and perhaps with a generally bluer color. Suppose this source, like the clouds, is inhomogeneously distributed in longitude (though it may or may not be causally connected with the clouds). It will then produce photometric variations that influence the observed light curve at each wavelength to an extent that depends on the ratio of the new spectrum to the photospheric spectrum at that wavelength. Hence, the new source will have little effect on the photometric light curve in the red continuum, but a very large effect in regions of strong alkali absorption because the photospheric emission there is so faint.

Under this supposition, the new source of flux must have been distributed similarly to the clouds on Feb 24 (though not identically: the light curve in the potassium band is not a perfect inverse of that in the red continuum), while on Feb 25 its distribution must have been offset from the clouds in such a way as to produce a phase shift in regions where the photosphere is faint (though again, we do not see a pure phase shift: there are changes in the light curve shape and amplitude also). If such a source of flux existed, it need not be causally linked to the clouds, and likely would not have anything to do with the neutral alkali abundances. In the next two sections, we consider astrophysical candidates for this source.

\subsection{Lightning} \label{sec:lightning}

Lightning is one possible source of bright emission with a spectrum very different from the photosphere. It has long been considered likely on brown dwarfs, due to their dynamic, high-energy clouds and the ubiquity of lightning in Solar System atmospheres \citep{Helling11a,Helling11b,Hodosan16}. Lighting could explain our Feb 24 data by producing faint blue continuum emission correlated with the dark clouds. Where the dominant red continuum of the brown dwarf photosphere is bright, this blue continuum would have negligible effect on the amplitude and phase. In regions where the red continuum is faint --- i.e., the alkali lines --- the blue continuum from the lightning would overwhelm the variations of the underlying photosphere and produce the observed amplitude reversal. Our Feb 25 results could be explained by supposing that only a subset of clouds is able to produce lightning, and then in the course of the general atmospheric disorganization that appeared to take place on Feb 25, this subset of clouds fell out of longitudinal alignment with the mean continuum obscuration, which perhaps by then was dominated by clouds unable to produce lightning.

This hypothesis implies extremely energetic lightning on Luhman 16B. It would remain a tiny fraction of the bolometric flux, but would still be orders of magnitude brighter than the lightning luminosities estimated by \citet{Hodosan16} for the extrasolar planet HD 189733b, which is a cloudy object of comparable temperature and radius to Luhman 16B (albeit with much lower mass). However, the best estimates of extrasolar lightning still rely on scaling from Solar System bodies (e.g., Jupiter), and are highly uncertain when extrapolated to hotter and more massive objects such as Luhman 16B. It is not impossible that lighting could be more common in the violently evolving silicate clouds of brown dwarfs than on any object in the Solar System.

The lightning hypothesis makes a further prediction that can be tested using our data. It predicts that the flux of Luhman 16B should extend into the extreme blue where no measurable photospheric flux is expected, and that the amplitude of variation there should be extremely large, approaching 100\%. Shortward of 6400\AA, the trace from Luhman 16 is too faint for us to deconvolve the separate objects, but we have already seen (Figure \ref{fig:synunres}) high-amplitude variations in unresolved photometry from 6000-6400\AA~that are correlated with those in the potassium line. Shortward of 6000\AA~the flux becomes too faint for time-series photometry, but by stacking all of our spectra we have managed to extract meaningful unresolved spectra at wavelengths down to 5250\AA~(Figure \ref{fig:resspec}).

There is less flux in the extreme blue than the lightning hypothesis would seem to predict. Our unresolved synthetic photometry (Figure \ref{fig:synunres}) shows a variability amplitude of 7\% in the potassium channel (7530--7830\AA). The average specific flux in this regime is $8.19\times 10^{-20} \mathrm{W}\mathrm{m}^{-2}\AA^{-1}$. If the amplitude reversal is caused by lightning, the specific flux from the lightning must be at least 7\% of this value, or $5.73\times 10^{-21} \mathrm{W}\mathrm{m}^{-2}\AA^{-1}$, a flux level indicated by the dashed green line in Figure \ref{fig:resspec}. If the lightning is producing a blue continuum, it should have similar or greater specific flux out to 5000\AA~ and shorter wavelengths. However, Figure \ref{fig:resspec} shows that the total flux drops below this value from 5400-6100\AA. This means the \textit{total} measured flux for Luhman 16 is less than we would predict for the flux from \textit{lightning alone} on \textit{Luhman 16B alone}. Making matters even worse, the very faint spectral trace in the 5250-5400\AA~ regime on the stacked images clearly aligns with Luhman 16A rather than Luhman 16B: there is no evidence that any of the flux at wavelengths shorter than 6000\AA~ comes from Luhman 16B.

This lack of blue flux does not explain away the huge variations we have detected in the 6000-6400\AA~band. These variations seem (based on their correlation with the potassium band) to come from Luhman 16B. Perhaps the lightning simply has an unexpectedly red spectrum, with little flux blueward of 6000\AA. To evaluate this possibility, we offer a brief analysis of the energetics. In unresolved synthetic photometry, we measure an amplitude reversal in both the potassium channel (7530--7830\AA) and the extreme blue (6000--6400\AA). The integrated fluxes in these regimes are $2.46 \times 10^{-17} \mathrm{W}\mathrm{m}^{-2}$ and $5.27 \times 10^{-18} \mathrm{W}\mathrm{m}^{-2}$, respectively, and the amplitudes are 7\% and 15\%. Multiplying the integrated fluxes by the amplitudes and adding, we set a spectrum-independent lower limit of $2.52 \times 10^{-18} \mathrm{W}\mathrm{m}^{-2}$ on the lightning flux from Luhman 16B, noting that in the case of realistic spectrum that would not have all its flux in the two regions where we can measure it, the total integrated flux would be larger. This spectrum-independent lower limit on the flux corresponds to a lightning luminosity of $1.20 \times 10^{17} \mathrm{W}$ for Luhman 16B. For comparison, \citet{Hodosan16} scale from measurements made on Jupiter and Saturn to estimate the total lightning energy released on the giant extrasolar planet HD 189733b in the course of a 1.89 hr transit to be $10^{17}$ J, of which a maximum of 10\% could be released in optical light. This produces a luminosity of $1.5 \times 10^{12} \mathrm{W}$, which is 80,000 times less than the absolute minimum lightning flux required to explain our Luhman 16 results in terms of lightning. Adopting any realistic spectrum for the lightning would raise this number by a factor of several.

Finally, the amplitude reversal in the potassium line itself is probably inconsistent with lightning. Lightning would be down in the clouds, with plenty of neutral alkali atoms still above it to wipe out the flux. The alternative would be something analogous to the `sprites' observed on Earth, which extend very far above the clouds --- but presumably such emission would involve only a tiny fraction of the total energy released in each lightning bolt, making the luminosity requirements even harder to meet.

The lack of blue flux from Luhman 16B; the amplitude reversal in the potassium line where lightning would probably be absorbed; and the implausible considerations in terms of total energy all suggest that lightning is not the explanation for our observations.

\subsection{Auroral Emission} \label{sec:aurora}

Aurora-like magnetic activity has been observed for some brown dwarfs \citep{Hallinan07,Berger09}, and would naturally appear above the alkali absorption in the atmosphere of Luhman 16B. It might explain our data by filling in spectral regions where the photosphere is faint. However, aurorae are not usually expected to produce continuum emission, and our results do not appear consistent with line emission. High-altitude fluorescence in the potassium line, for example, would create a highly localized signature at the core of the line, which would not match the broad effects we see. If aurora did create a hot continuum, it would be inconsistent with the lack of flux in the extreme blue that we have already invoked to rule out lightning. Finally, \citet{Osten2015} have made extremely sensitive radio observations of Luhman 16, ruling out auroral activity except at a very low level.

\section{The Complex Evolution of Clouds} \label{sec:interp}

\subsection{Variable Alkali Abundances} \label{sec:alk}

On February 24, we observed an amplitude reversal in the 7682\AA~potassium doublet of Luhman 16B relative to its red continuum, while on the same night in infrared lines of the same element, \citet{comp} observed an amplitude increase that remained in phase with the red continuum. This apparently surprising discrepancy can likely be resolved by recognizing that because the optical lines of the alkali metals are far stronger than the near-infrared lines, the two wavelength regimes do not probe the same layers of the atmosphere. The optical lines we have observed probe far higher altitudes than the infrared lines measured by \citet{comp}, and the neutral alkali abundances could be different. In considering this, we depart from the simple conceptual model of Section \ref{sec:expectation}, in which longitudinal variations in cloud cover were the only important cause of variability. Now, we picture a more complex and realistic atmosphere in which clouds, gas composition, and temperature profile are mutually interrelated; all three are longitudinally inhomogeneous; and the dominant cause of photometric variability could change as a function of wavelength. Specifically, we suggest that longitudinal inhomogeneity of obscuring clouds dominates the variability in the red continuum, but that inhomogeneous abundance of the neutral alkalis (influenced by clouds) becomes an important source of variability in the alkali lines.

\citet{comp} note that their observations could be explained by a greater abundance of potassium above the dark clouds in Luhman 16B relative to the clear regions (though they do not ultimately favor this explanation). Our data could similarly be explained by a {\em lower} abundance of potassium over the dark clouds -- but at a much higher altitude. Hence, both data sets could be explained if clouds in Luhman 16B affect the temperature and chemistry of the atmosphere above them in a way that increases the abundance of neutral alkali metals at low altitude (just above the clouds), while reducing it at high altitude. Proposing a detailed mechanism to produce such effects is beyond the scope of this work, but we note that in T dwarf atmospheres, the neutral alkali metals are in a chemical equilibrium with their chlorides \citep[e.g.,][]{Burrows2002,comp}, and hence their abundances are sensitive to local conditions that could foster either the formation or dissociation of NaCl and KCl. We speculate that such conditions could involve not only cloud-driven changes in the vertical temperature profile, but also horizontal or vertical transport of gas with temporarily non-equilibrium chemical abundances --- and additionally that solid particles in the clouds might act as catalysts for either the formation or dissociation of the alkali chlorides. 

We can also interpret the Feb 25 results under this basic framework. On Feb 25, the decrease in overall amplitude indicates that the atmosphere of Luhman 16B was becoming less organized. As part of this disorganization, the alkali-poor masses of the high atmosphere might have fallen out of longitudinal alignment with the mean obscuration of the red continuum, causing the observed phase shift. We would then predict that at any time when the atmosphere of Luhman 16 is strongly organized into cloudy and less-cloudy regions, and hence the amplitude of variation is large, the well-organized clouds will again deplete the alkali gas at high altitudes above them (whatever the physical mechanism by which they do this), and an amplitude-reversal will again be observed in the Na D and potassium 7682\AA~ lines. It appears that such conditions were already observed by \citet{Biller2013}, though the inverted amplitudes were diluted by their broadband $r'$ and $i'$ filters. Our prediction could be further tested by photometrically monitoring Luhman 16 using customized filters to deliberately probe the sodium and potassium lines, and comparing the results with simultaneous monitoring in the red continuum. Such monitoring could be an interesting tool for studying not only Luhman 16 but also other optically variable brown dwarfs.

\subsection{Pressure-Dependent Light curve Changes: Not a Simple Phase-Shift} \label{sec:pressure}

Several previous investigations have found significant wavelength-dependent differences in simultaneously-measured brown dwarf light curves, and have argued convincingly that these are due to the differing pressure levels probed by the various wavelengths \citep[e.g.,][]{Buenzli2012,Yang2016,Biller2018}. Our results clearly fall into the same broad category. However, we do not take the further step, made in some previous works, of characterizing our results as indicating a pressure-dependent phase shift --- even though it is tempting to interpret our Feb 25 data this way. We avoid doing this because we believe the concept of phase shifts smoothly varying with pressure is not astrophysically useful. The pressure-dependent phase shifts that have been suggested in some past works cannot be reconciled with any coherent physical picture of the cloud structures on the brown dwarf because the range of phases is too large. For example, a cloud band with a longitude dependent altitude (i.e., a cloud configured like a ramp) would produce a pressure-dependent phase shift, but the full range of the phase variation could not be more than a few degrees unless the ramp was very shallow. Even if it were an {\em extremely} shallow ramp that encircled the entire brown dwarf, the range of phases would have to be less than 180 degrees --- unless there were a contrast reversal, which would produce a 180-degree phase shift (i.e., an amplitude inversion).

Such amplitude inversions at different wavelengths (i.e., pressure levels) have been observed by, e.g., \citet{Buenzli2012}, \citet{Biller2018}, and ourselves --- but where a 180 degree phase shift exists between wavelengths that probe different pressure levels, there is no physical reason to expect {\em any} range of pressure levels to fill in a smooth variation of phase shift from zero to 180 degrees. If anything, the quantity that might vary smoothly with pressure in such a case would be the amplitude. Our data appear to show something like this for Luhman 16B on February 24, though on February 25 we see something more like a phase shift. However, the lack of a coherent physical interpretation for (wide-ranging) pressure dependent phase shifts leads us to regard the observations as simply showing large pressure-dependent {\em changes} in the light curves observed by ourselves and others, not simple phase-shifts. This broader view is better able to accommodate the observations reported by \citet{Yang2016}, which in several cases show large pressure-dependent changes in light curve shape that are very different from pure phase shifts --- as well as our own observations of pressure-dependent changes that manifest primarily as amplitude variations on February 24 and phase variations on February 25, with additional shape changes that cannot be simply captured with either amplitude or phase.

\subsection{Complex Wavelength-Dependent Changes: The Expected Norm?} \label{sec:norm}

In Section \ref{sec:alk} we put forth our best attempt to explain the observed variability of Luhman 16B on February 24 and 25 in terms of effects connected with a single dominant cloud system that was strongly organized on Feb 24 and became less organized on Feb 25. Following the principle of Occam's Razor, we avoid proposing {\em multiple} major cloud systems where a single system might explain the data. Nevertheless it is worth noting that on Jupiter --- the best Solar System analog to a brown dwarf --- multiple cloud systems do exist (i.e., in different ranges of latitude), and very large pressure-dependent light curve changes are seen in part because different cloud systems dominate the rotational light curve at different wavelengths \citep{Ge2019}. For example, in the 8890\AA~ methane band, \citet{Ge2019} find Jupiter's rotational light curve to be dominated by the Great Red Spot, while in the 5$\mu$m atmospheric window it is dominated by the North Equatorial Belt. The difference between the variations seen in the two wavelengths cannot be well characterized either by an amplitude inversion or by a phase shift: the 8890\AA~ light curve is double-peaked (though asymmetrical), while the 5$\mu$m light curve is single-peaked (though not sinusoidal). At yet a third wavelength (8.80$\mu$m), \citet{Ge2019} find a double-peaked light curve so symmetrical that if it were the only information we had about Jupiter, we would certainly claim its rotational period to be half the true value. Since there is no reason to think brown dwarf atmospheres are simpler than that of Jupiter, complex wavelength-dependent changes not fitting any simple model should perhaps be our default expectation.

\subsection{Rapid Time-Evolution of Light curves: Luhman 16B and Neptune} \label{sec:Neptune}

Both from our data and many other results \citep[e.g.,][]{Gillon2013,Apai2021}, the light curve of Luhman 16B is known to evolve very rapidly with time, changing shape and amplitude (by a factor of two or more) on the timescale of one or two rotations. This suggests violent winds driving global changes in cloudcover, but it is interesting to note that the optical light curve of Neptune \citep{Simon2016} exhibits similarly large fractional changes on timescales comparable to its rotation period. Neptune is far colder, less massive, and slower-rotating ($P\sim 17$ hr) relative to Luhman 16B, and its typical photometric amplitude ($\sim$1\%) is about ten times smaller. Hubble Space Telescope observations suggest its variations are mostly due to relatively small, high-contrast clouds \citep{Simon2016}. A light curve produced by cloud features much smaller than a hemisphere could change dramatically without extremely fast winds or global atmospheric upheaval. On the other hand, unless the cloud features responsible for the variations of Luhman 16B have extremely high contrast, its much larger photometric amplitude implies the cloud patches must be larger relative to the object's radius than those of Neptune --- a scenario supported by the \citet{Crossfield2014} finding of large dark and bright regions with only moderate brightness contrast. The large size of these regions relative to the clouds of Neptune suggests that more violent weather is required to explain rapid light curve evolution in Luhman 16B. Such transformations could involve either vertical or horizontal transport of large masses of gas, perhaps faster than the alkali-chloride chemical equilibrium could adjust to the changing conditions. Transport of this type might explain the apparent transformation of the alkali effect from an amplitude inversion to a phase shift between Feb 24 and Feb 25.

\section{Conclusion} \label{sec:conc}

We have monitored the binary brown dwarf Luhman 16 using Gemini/GMOS for two nights (UT 2014 Feb 24 and 25), probing wavelengths from 6000--10500\AA~ with precise spectrophotometry, and obtaining a spectrum with measurable flux all the way down to 5250\AA. Wide, 5-arcsecond slits prevented seeing-dependent slit losses from contaminating our photometry, while the GMOS nod-and-shuffle mode enables clean subtraction of the sky background despite severe interference fringing at long wavelengths. Careful forward-modeling of spectral trace functions in the presence of differential chromatic refraction enables us to deconvolve the spectra of Luhman 16 A and B despite their projected separation of only 1.13 arcsec.

Consistent with previous results, we find that in the red continuum (e.g., 8000-10000\AA), Luhman 16B is variable with a period of about 5 hr and amplitude ranging from 13\% on Feb 24 to 7\% on Feb 25. Luhman 16A shows smaller variations, and on Feb 24 showed no variability down to a measurement uncertainty of about 1\%. In wavelength regimes affected by the hugely-broadened neutral alkali doublets (sodium 5893\AA~ and potassium 7682\AA), the photometric behavior of Luhman 16B diverges strongly from its red continuum behavior, with the largest differences at the wavelengths of strongest alkali absorption (i.e., the cores of the lines and their inner wings). In these spectral regions we see an amplitude inversion relative to the red continuum on Feb 24, and a phase-shift on Feb 25. The inversion and phase shift are not pure effects but are simply the dominant changes observed on the respective nights. In the Feb 24 data, we estimate a variability amplitude of at least 30\% in the core of the potassium line for Luhman 16B, more than twice as large as the (anticorrelated) variations in the red continuum.

Previously published photometric monitoring of Luhman 16B also suggests amplitude inversions in the optical alkali lines relative to the red continuum. In data from April 2013, \citet{Biller2013} noted amplitude inversions in the $r'$ and $i'$ bands (which contain the 5893\AA~sodium and the 7682\AA~potassium doublets, respectively) relative to the $z'$, $J$, and $H$ bands. In observations made November 2014, \citet{Buenzli2015b} saw a strong decrease in amplitude in the red wing of the 7682\AA~potassium doublet, though their data did not reach short enough wavelengths to see the amplitude inversion that probably existed in the core of the line. Although our Feb 25 data show that the atmosphere of Luhman 16B is not always in a state that produces the alkali-correlated amplitude inversion we saw on Feb 24, these previously published studies indicate that finding the object in a such a state is not unusual.

While we observed an amplitude inversion in the optical potassium doublet on Feb 24, in simultaneous observations on the same night \citet{comp} saw positively correlated {\em increased} amplitude as compared to the red continuum in Luhman 16B's {\em infrared} potassium absorption lines (1.173 $\mu$m and 1.248 $\mu$m). The resolution of this apparent inconsistency could lie in the fact that the optical lines we have probed form at very high altitudes in the atmosphere of Luhman 16B, while the near-infrared lines probed by \citet{comp} form somewhat deeper. Both our observations and those of \citet{comp} could be explained by reduced abundances of neutral sodium and potassium at altitudes very far above the clouds, coupled with increased neutral potassium at lower altitudes nearer the cloud tops. 

Altitude-dependent variations in alkali abundances are rendered more plausible by the fact that the alkalis are believed to be in chemical equilibrium with their chlorides in brown dwarf atmospheres \citep{Burrows2002}, and changing local conditions could tilt the equilibrium and greatly change the abundances. Though not intuitive, it seems plausible that clouds could cause an increase in the abundance of neutral alkali metals immediately above them and a decrease at higher altitudes --- by changing the atmospheric temperature profile and/or catalyzing the formation or dissociation of the alkali chlorides. The higher-amplitude red continuum variations we measure on Feb 24 suggest a large, well-organized cloud system that might have created strong, well-aligned variations in alkali abundance. The decreased variability amplitude on Feb 25 suggests the clouds were becoming less organized, and the phase-shift that is the dominant alkali effect on that night could indicate the most chemically influenced regions of the atmosphere were drifting out of alignment with the clouds and perhaps were in flux towards a new chemical equilibrium.

While do not claim that this tentative proposal is the only one that could fit our data, we believe that plausible cloud-based explanations exist for all of Luhman 16's complex behavior: hence, there is no need to invoke non-photospheric processes such as lightning or aurorae. Furthermore, our analysis suggests lightning and aurorae are problematic as explanations both because they are unlikely to produce the observed spectral behavior, and because the implied energy budgets for either Luhman 16B's lightning storms or its magnetosphere appear implausibly high.

We have hypothesized a causal link between large, organized cloud masses on Luhman 16B and atmospheric conditions that produce an amplitude inversion in the optical alkali lines. Further photometric monitoring of Luhman 16B in customized filters targeting the 7682\AA~potassium and 5893\AA~sodium lines (simultaneously with the red continuum) can test this idea and address further questions about the object's weather. Are large-amplitude variations in the near infrared and the red continuum always accompanied by an amplitude inversion in the alkali lines, as our hypothesis suggests? Does the inversion always fade when the atmosphere transitions to a less ordered state and the red continuum amplitude drops? What can the intermediate states tell us about wind shear and/or chemical evolution in the clouds of Luhman 16B? The customized alkali filters required for such investigations could be useful tools for studying the atmospheres of variable brown dwarfs in general --- and the needed observations could be obtained using smaller telescopes on which the time required for extensive monitoring programs might be more readily available. 

While the detailed causes remain uncertain, our observations of Luhman 16B reveal a dynamic atmosphere with indications of unstable chemical equilibria influenced by altitude and cloudcover. The object's high temperature and strong gravity evidently enable global changes in cloudcover that go well beyond what is seen in the colder atmospheres of the giant planets in our own solar system.

\section{Acknowledgments} 
This research is based on observations obtained at the international Gemini Observatory, a program of NSF's NOIRLab, which is managed by the Association of Universities for Research in Astronomy (AURA) under a cooperative agreement with the National Science Foundation on behalf of the Gemini Observatory partnership: the National Science Foundation (United States), National Research Council (Canada), Agencia Nacional de Investigaci\'{o}n y Desarrollo (Chile), Ministerio de Ciencia, Tecnolog\'{i}a e Innovaci\'{o}n (Argentina), Minist\'{e}rio da Ci\^{e}ncia, Tecnologia, Inova\c{c}\~{o}es e Comunica\c{c}\~{o}es (Brazil), and Korea Astronomy and Space Science Institute (Republic of Korea). These observations were obtained under Gemini Program ID GN-2014A-C-1.

This work relies also on observations from the TRAPPIST telescope. TRAPPIST is funded by the Belgian Fund for Scientific Research (Fond National de la Recherche Scientifique, FNRS) under the grant FRFC 2.5.594.09.F, with the participation of the Swiss National Science Foundation (SNF). MG is F.R.S.-FNRS Senior Research Associate. The authors of this work thank E. Jehin (Universit\'{e} de Li\`{e}ge) for his crucial role in the operations of TRAPPIST at the time of the observations.

This research was  supported by NASA through the Spitzer Exploration Science Program {\it 
Weather on Other Worlds} (program GO 80179) and ADAP award NNX11AB18G.
This publication makes use of the SIMBAD online database,
operated at CDS, Strasbourg, France, and the VizieR online database (see \citet{vizier}).

RK acknowledges support by ANID --- Millennium Science Initiative Program --- ICN12 009 awarded to the Millennium Institute of Astrophysics (MAS).

This publication makes use of data products from the Two Micron All Sky Survey, 
which is a joint project of the University of Massachusetts and the Infrared Processing
and Analysis Center/California Institute of Technology, funded by the National Aeronautics
and Space Administration and the National Science Foundation.
We have also made extensive use of information and code from \citet{nrc}. 
We have used digitized images from the Palomar Sky Survey 
(available from \url{http://stdatu.stsci.edu/cgi-bin/dss\_form}),
 which were produced at the Space 
Telescope Science Institute under U.S. Government grant NAG W-2166. 
The images of these surveys are based on photographic data obtained 
using the Oschin Schmidt Telescope on Palomar Mountain and the UK Schmidt Telescope.

\facilities{Gemini South, TRAPPIST}

\end{document}